\documentclass{article}

\usepackage{PRIMEarxiv}

\usepackage[utf8]{inputenc} % allow utf-8 input
\usepackage[T1]{fontenc}    % use 8-bit T1 fonts
\usepackage{hyperref}       % hyperlinks
\usepackage{url}            % simple URL typesetting
\usepackage{booktabs}       % professional-quality tables
\usepackage{amsfonts}       % blackboard math symbols
\usepackage{nicefrac}       % compact symbols for 1/2, etc.
\usepackage{microtype}      % microtypography
\usepackage{lipsum}
\usepackage{fancyhdr}       % header
\usepackage{graphicx}       % graphics
\graphicspath{{media/}}     % organize your images and other figures under media/ folder
\usepackage{amsmath} 
\usepackage{multirow}

\usepackage[final]{changes}
\definechangesauthor[color=blue]{ac}
\definechangesauthor[color=blue]{pm}
\definechangesauthor[color=blue]{rt}

\newcommand{\myparagraph}[1]{\vspace{1mm}\noindent{\textit{#1}}}

\newcommand{\oo}{\textbf{o}}
\newcommand{\ft}[1]{{\textbf{#1}}}
\newcommand{\f}{\textbf{a}}

\newcommand{\horaug}{\alpha^{\rightarrow}}
\newcommand{\veraug}{\alpha^{\small\downarrow}}
\newcommand{\join}{\bowtie}
\newcommand{\union}{\uplus}

\newtheorem{example}{Example}[section]

\title{Supporting Better Insights of Data Science Pipelines with Fine-grained Provenance
}

\author{
  Adriane Chapman \\
  University of Southampton, UK\\
  \texttt{Adriane.Chapman@soton.ac.uk} \\
   \And
  Luca Lauro \\
  Universit\`a Roma Tre, Italy \\
  \texttt{luca.lauro@uniroma3.it} \\
   \And
  Paolo Missier \\
  Newcastle University, UK \\
  \texttt{lpaolo.missier@ncl.ac.uk} \\
   \And
  Riccardo Torlone \\
  Universit\`a Roma Tre, Italy \\
  \texttt{riccardo.torlone@uniroma3.it} \\
}

\begin{document}

\sloppy

\maketitle

\begin{abstract}
Successful data-driven science requires complex data engineering pipelines to clean, transform and alter data in preparation for machine learning, and robust results can only be achieved when each step in the pipeline can be justified, and its effect on the data  explained. 
In this framework, our aim is to provide data scientists with facilities to gain an in-depth understanding of how each step in the pipeline affects the data, from the raw input to training sets ready to be used for learning.
Starting from an extensible set of data preparation operators commonly used within a data science setting, in this work we present a provenance management infrastructure for generating, storing, and querying very granular accounts of data transformations, at the level of individual elements within datasets whenever possible.
Then, from ~the formal definition of a core set of data science preprocessing operators, we derive a \textit{provenance semantics} embodied by a collection of templates expressed in PROV, a standard model for data provenance.   Using those templates as a reference, our provenance generation algorithm generalises to any operator with observable input/output pairs. We provide 
a prototype implementation of an application-level provenance capture library to produce, in a semi-automatic way, complete provenance documents that account for the entire pipeline.
We report on that reference implementations ability to capture provenance in real ML benchmark pipelines and over TCP-DI synthetic data. We finally show how the collected provenance can be used to answer a suite of provenance benchmark queries that underpin some common pipeline inspection questions, as expressed on the Data  Science  Stack  Exchange. 
\end{abstract}

\keywords{Provenance, Data Science, Data Preparation, Preprocessing}

\section{Introduction} \label{sec:introduction}

Dataset selection and data wrangling pipelines are integral to applied Data Science workflows.  These typically  culminate in the generation of predictive models for a broad range of data types and application domains through training.
A number of critical choices are made when these pipelines are designed, starting with the choice of which datasets to include or exclude, how these should be merged \cite{10.1145/3035918.3058730},
and which transformations are required to produce a viable training set, given a choice of target learning algorithms. 
The main intended consequence of these transformation pipelines is to optimise the predictive performance and generalisation characteristics of the models that are derived from the ground data. 
There are however also unintended consequences, as these transformations alter the representation of the domain that  the learning algorithms generalise from, and they may remove or inadvertently introduce new bias in the data \cite{ghorbani2019data}. 
In turn, this may reflect on non-performance properties of the models, such as their \textit{fairness}.
The term, formally defined in terms of statistical properties of the model's predictions~\cite{narayanan2018translation}, broadly refers to the capability of a model to ensure that its predictions are not affected by an individual belonging to one of the groups defined by some sensitive attribute(s), such as sex, ethnicity, income band, etc.

\myparagraph{Motivation.}
Models that are provably fair are also perceived as more trustworthy, an important feature at a time when machine learning models are increasingly used to support and complement human expert judgment,  in areas where decisions have consequences on individuals as well as on businesses. Substantial recent research has produced techniques for explanation using: counterfactuals~\cite{mothilal2019explaining}, local explanations~\cite{ribeiro2016should}, data~\cite{Lakkaraju2017}  and meta-models~\cite{alaa2019demystifying}. 
While these techniques focus primarily on the model itself, relatively little work has been done into trying to explain models in terms of the transformations that occur \textit{before} the data is used for learning.
The ultimate goal of this work is to enable explanations of the effect of each transformation in a pre-processing pipeline on the data that is ultimately fed into a model \cite{grafberger2022data}. As an initial step in this direction, we have developed a formal model and practical techniques for recording data derivations at the level of the atomic elements in the dataset, for a general class of data transformation operators. These derivations are a form of data provenance and are expressed using the PROV data model~\cite{w3c-prov-dm}, a standard and widely adopted ontology. Data derivations form a corpus of graph-structured metadata that can be queried as a preliminary step to support user questions about model properties.

\myparagraph{Problem scope.}
\added[id=pm]{In this paper we focus on data transformations that are commonly found in data science processing pipelines and across application domains, and we further limit the scope to structured tabular data.}\footnote{However, we are not going to consider more specialised data pre-processing steps that may apply to data types such as video, audio, images, etc.} 
These steps have been systematically enumerated in multiple reviews (see e.g.~\cite{Garcia2016,Mirza2019})  and include, among others: feature selection, engineering of new features; imputation  of missing values, or \textit{listwise deletion} (excluding an entire record if data is missing on any variable for that record); downsampling or upsampling of data subsets in order to achieve better balance, typically on the class labels (for classification tasks) or on the distribution of the outcome variable (for regression tasks); outlier detection and removal; smoothing and normalisation; de-duplication, as well as steps that preserve the original information but are required by some algorithms, such as ``one-hot'' encoding of categorical variables.
A complex pipeline may include some or all of these steps, and different techniques, algorithms, and choice of algorithm-specific parameters may be available for each of them. These are often grounded in established literature but variations can be created by data scientists to suit specific needs. 
In this work we consider the space of all configured pipelines that can potentially be composed out of these operators.

\added[id=pm]{Regarding the data that these operate on, we focus on structured two-dimensional tabular data, namely dataframes, which are commonly supported by R and python as well as by a dedicated Spark API, and excluding tensors and multidimensional matrices.
While this is done to simplify our proof-of-concept implementation, we observe that considering higher-dimension tabular structures has practical implications as it increases the complexity of the derivations from input to output elements, however the underpinning provenance templates are fundamentally the same.}

\myparagraph{Overview of the approach.}
\label{sec:overview}
Firstly, we provide  a formalisation and categorisation of a core set of these operators.
Then, with each class of those operators, we associate a \textit{provenance template} that describes the effect on the data of each operator in the class at the appropriate level of detail, i.e., on individual data elements, columns, rows, or collections of those. 
By mapping operators to these fundamental templates, we are then able to identify the transformation type based on observation of the operator's input and outputs alone. By abstracting to this level, we can automatically create the appropriate provenance for an operator in a data science pipeline if it follows the pre-identified input-output patterns, even if the operator itself has never been seen before.
% Effectively, the provenance patterns play a similar role to that of \textit{provenance polynomials}~\cite{green2007}, i.e., annotations that are associated to relational algebra operators to describe the fine-grained provenance of the result of relational as well as linear algebra operators \cite{WuTD20,YanTI16}.
%
%  We then associate a provenance function $\mathit{pf_{\op}}()$ to each operator $\op$, which generates a provenance document $\mathit{pf_{\op}}(D)$ when a dataset $D$ is processed using $\op$. The provenance document is an instance of the pattern associated with $\op$.
% Provenance functions are then implemented as part of a python library. 
%
% Collecting in a semi-automatic way all the provenance documents from each operator's execution results in a seamless, end-to-end provenance document that contains the detailed history of each dataset element in the final training set, including their creation (e.g. as a new derived feature), transformation (value imputation, for example) and possibly deletion (e.g., by feature selection, removal of null values).

\myparagraph{Contributions.}
Our contributions can be summarised as follows.
\begin{itemize}%\itemsep0pt \parskip0pt \parsep0pt
\item A formalisation and categorisation of a core set of operators for data reduction, augmentation, transformation, and fusion that move beyond the relational algebra (Section~\ref{sec:operators}), showing how common data pre-processing pipelines can be expressed as a composition of these operators.
\item The semantics of the provenance that is generated for white-box transformations, as reduced to the core set of operators (Section ~\ref{sec:provenance}).
\item A method for capturing the provenance of a pipeline, based on \textit{observing the changes to the data}, not the operator that was applied (Section~\ref{sec:algorithms}).
%, based on a reduced collection of provenance-generating function, one for each core operators, that are invoked alongside the execution of the steps of the pipeline;
\item An application-level provenance capture facility for Python, underpinned by the formal model, that (i) identifies the operation under execution to capture its provenance and (ii) is backed by a Neo4J database used as a provenance store (Section~\ref{sec:architecture}). This new approach almost entirely removes the older requirement for pipeline designers to programmatically ``drive'' provenance generation, making most of the process transparent;
%\item A validation of the query capabilities of the resulting granular provenance, using a collection of machine learning datasets using real data pre-processing pipelines, to show that using the resulting provenance we can successfully answer a suite of provenance queries from the Data Science Stack Exchange\footnote{\url{https://datascience.stackexchange.com}} (Section \ref{sec:evalrealperf}).
\item \added[id=ac]{Using a reference implementation, we report on: (i) the impact of adding provenance capture to real-world pipelines (Section \ref{sec:evalrealperf}), (ii) the ability to capture provenance in real ML benchmark pipelines and over TCP-DI synthetic data  \cite{tpcdi} (Section \ref{sec:tpcdi}), (iii) a use case analysis showing that provenance queries can provide support to data scientists in the development of real-world machine-learning pipelines (Section \ref{sec:dsseanalysis})
(iv) how data provenance collected with our approach can be inspected through user-friendly interfaces (Section \ref{sec:provexp}), and (v) a comparison to other similar provenance capture systems (Section \ref{sec:comparison}).} \deleted[id=ac]{A scalability analysis showing that, while the overall provenance document can be arbitrarily large, it is created incrementally in a persistent data store, making the entire process scalable  in the number of operators (Section \ref{sec:tpcdi}). We  run extensive experiments on a synthetic TPC-DI dataset at multiple scales \cite{tpcdi}, and report on the time and space overhead of using the provenance functions.}
\end{itemize}

% While the main purpose of these manipulations is to maximise the model's predictive performance, they may have other consequences, for instance by inadvertently introducing bias in the dataset, which in turn may affect non-performance properties of the model.

% Transparency, accountability, and fairness are three such properties. These are  important as they contribute to making the model trustworthy, at a time when predictive machine learning models are increasingly used to support and complement human expert judgment, in areas where decisions have consequences on individuals as well as on businesses. 
% The ability to understand such properties, and how they affect the model generation process, is therefore an important practical concern.
% This recognition has spurred research to add explainability features to ML models~\cite{alaa2019demystifying,Lakkaraju2017}, and to ensure that models are fair to those who are affected by the predictions they make.

% The main hypothesis explored in this work is that in order to enforce these properties on the model, it is useful to understand the transformations that occur in the pipeline, where the data used to learn the model is prepared.
% To explore this, we have developed a formal model and practical techniques for recording data derivations at the level of the atomic elements in the dataset, that is, by recording the final training set's fine-grained provenance through its transformations in the pipeline.

\section{Models and Problem statement}\label{sec:context}

\subsection{Data model}
\label{sec:datamodel}

The data collected for ML tasks are usually represented as tables or statistical data matrices in which columns represent specific features of a phenomenon being observed, and rows are records of data for those features describing observations of the phenomenon. To capture both formats, we will refer to these generically as \emph{datasets}, similar in spirit to notions of ordered relations~\cite{Cheung15} and dataframes~\cite{PetersohnMLMXMG20}.

A \textit{(dataset) schema} $S$ is an array of distinct names called \emph{features} (or \textit{attributes})
$S=
    [
    \f_{1}, \dots, \f_{n}
    ].
$
Each feature is associated with a domain of atomic values (such as numbers, strings, and timestamps). With a little abuse of notation, hereinafter we will compare schemas using
set containment over their features.
A \emph{dataset} $D$ over a schema     $S=[\f_{1}, \dots, \f_{n}]$
is an ordered collection of \emph{rows} (or \emph{records}) of the form:
$i:(d_{i1}, \dots, d_{in})$
where $i$ is the unique \emph{index} of the row and each element $d_{ij}$  (for $1\leq j\leq n$) is either a value in the domain of the feature $\f_j$ or the special symbol $\bot$, denoting a missing value. Row indexes can be implemented in different ways (e.g., with RID annotations~\cite{psallidas2018smoke}). We only assume here that a row of any dataset can be uniquely identified.

Given a dataset $D$ over a schema  $S$ we denote by $D_{i\f}$ the value for the feature $\f$ of $S$ occurring in the $i$-th row of $D$. We also denote by $D_{i\ast}$ the $i$-th row of $D$, and by $D_{\ast\f}$ the column of $D$ associated with the feature $\f$ of $S$.

\begin{example}\label{ex:toyexample}
A possible dataset  $D$ over the schema  
$S=[
 \ft{CId}, \ft{Gender}, \ft{Age}, \ft{Zip}
]$
is as follows:
\[%\small
\begin{array}{|c|cccc|}
\hline
\multicolumn{1}{|c|}{} & \ft{CId}& \ft{Gender} & \ft{Age}& \ft{Zip}\\
\hline
    %\cline{2-5}
    1\ & 113 & F & 24 & 98567\\ 
    %\cline{2-5}
    2\ & 241 & M & 28 & \bot\\ 
    %\cline{2-5}
    3\ & 375 & C & \bot & 32768\\
    %\cline{2-5}
    4\ & 578 & F & 44 & 32768\\
    \hline
\end{array}
\]
$D_{\ast\ft{Age}}$ and $D_{2\ast}$ denote the third column and the second row of $D$, respectively.
\end{example}

\added[id=pm]{Note that, as mentioned in the introduction, in this work we focus on dataframes, which are described by a schema. Extensions to tensors and multidimensional matrices are left for future work. }

\subsection{Data manipulation model}
\label{sec:dataman}

%Building accurate predictive models or extracting useful patterns from large datasets requires careful data preparation before machine learning algorithms and other data analytics tools can be employed successfully.
%The terms data wrangling, or munging, are commonly used to describe the collection and typical sequence of data preparation steps, ranging from data source discovery to mapping and integration, and multiple types of transformation. 
 
%Extracting value from data using machine learning algorithms is a complex, multi-step process that  Machine learning algorithms generate 
%In this paper we focus on the data transformation steps 
%Building a Machine Learning model is a multistep process that starts from one of more datasets and typically involves a number of data transformation steps, aimed at preparing 

%Machine learning (ML) helps in automatically finding complex and potentially useful patterns in data. These patterns are condensed in an ML model that can then be used on new data points. . 

\myparagraph{A general classification.}

As part of this work, we analyzed several packages that allow users to build data pre-processing pipelines. Table \ref{tab:mloperators} contains an example overview of the available operators from the ML pipeline building tool Orange \cite{JMLR:demsar13a} and the popular SciKit packages \cite{scikit-learn}. 
% \rtnote{Changing the order of columns in Table \ref{tab:mloperators} could be helpful here to follow the discussion: Orange, Category, Operator, Implementation}
As indicated on the left-hand side of the table, all of them can be classified into four main classes, according to the type of manipulation done on the input dataset(s) over a schema $S$:

\begin{table*}\footnotesize%\small
\caption{Typical operations in ML pipelines of data preparation from Orange \protect\cite{demvsar2013orange} and Scikit-Learn \protect\cite{scikit-learn}.}
\label{tab:mloperators}
\centering
\begin{tabular}{|c|c|c|c|c|}
\hline
\textbf{Orange3 Ex.} & \textbf{Scikit\-Learn Ex.}
& \textbf{Category} & \textbf{Operator} & \textbf{Implementation}  \\
\hline \hline
Feature Statistics & Feature\_selection & \multirow{5}{8em}{\centering Data reduction} & Feature Selection  & $\pi_C$ \\
\cline{1-2} \cline{4-5}
Select Data by Index & Dataframe op. & & Instance Selection & $\sigma_C$  \\
\cline{1-2} \cline{4-5}
Select Columns & Feature\_selection & & Drop Columns & $\pi_C$ \\
\cline{1-2} \cline{4-5}
Select Rows & Dataframe op. & & Drop Rows & $\sigma_C$ \\
\cline{1-2} \cline{4-5}
Data Sampler & Imbalanced-learn & & Undersampling & $\sigma_C$ \\
 \hline
Impute & SimpleImputer  & \multirow{5}{8em}{\centering Data transformation} & Imputation & $\tau_{f(X)}$ \\
\cline{1-2} \cline{4-5}
Apply Domain & FunctionTransformer & & Value Transformation & $\tau_{f(X)}$ \\
\cline{1-2} \cline{4-5}
Edit Domain & Binarizer & & Binarization & $\tau_{f(X)}$ \\
\cline{1-2} \cline{4-5}
Preprocess & Normalizer & & Normalization & $\tau_{f(X)}$ \\
\cline{1-2} \cline{4-5}
Discretize & KBinDiscretizer & & Discretization & $\tau_{f(X)}$ \\
 \hline
Feature Constructor & FunctionTransformer & \multirow{5}{8em}{\centering Data augmentation} & Space Transformation & $\pi_Z\circ\horaug_{f(X): Y}$ \\
\cline{1-2} \cline{4-5}
Create Class & FunctionTransformer  & & Instance Generation & $\veraug_{X:f(Y)}$ \\
\cline{1-2} \cline{4-5}
Data Sampler & Imbalanced-learn & & Oversampling & $\veraug_{X:f(X)}$ \\
\cline{1-2} \cline{4-5}
Corpus & Label Encoder & & String Indexer & $\horaug_{f(X): Y}$ \\
\cline{1-2} \cline{4-5}
Preprocess & OneHotEncoder & & One-Hot Encoder &  $\horaug_{f(X): Y}$ \\
\hline
Merge & Hstack  & \multirow{2}{8em}{\centering Data fusion} & Join & $\join^t_{C}$ \\
\cline{1-2} \cline{4-5}
Concatenate & Vstack & & Append & $\union$ \\
\hline
\end{tabular}

\end{table*}

\begin{itemize}\itemsep0pt \parskip0pt \parsep0pt
    \item Data reductions: operations that take as input a dataset $D$ on a schema $S$ and reduce the size of $D$ by eliminating rows (without changing $S$) or columns (changing $S$ to $S'\subset S$) from $D$;
    \item Data augmentations: operations that take as input a dataset $D$ on a schema $S$ and increase the size of $D$ by adding rows (without changing $S$) or columns (changing $S$ to $S'\supset S$) to $D$;
    \item Data transformations: operations that take as input a dataset $D$ on a schema $S$ and, by applying suitable functions, transform (some of) the elements in $D$ without changing its size or its schema (up to possible changes to the domain of the involved features of $S$)
    \item Data fusions: operations that take as input two datasets $D_1$ and $D_2$ on schema $S_1$ and  $S_2$ respectively and combine them into a new dataset $D$ on a schema $S$ involving the features of $S_1$ and  $S_2$.\label{ch:domain}
\end{itemize}

We now introduce a number of basic operators of data manipulation over datasets belonging to one of the above classes of data manipulations, as indicated in the right-hand side of Table \ref{tab:mloperators}.
This approach is in line with the observation that most of the operations of current data exploration packages rely on a rather small subset of operators~\cite{PetersohnMLMXMG20}. \label{small}

\myparagraph{Data reductions.} Two basic data reduction operators are defined over datasets. They are simple extensions of two well-known relational operators.
\begin{description}
    \item [$\pi_{C}$:] the \textit{(conditional) projection} of $D$ on a set of features of $S$ that satisfy a boolean condition $C$ over $S$, denoted by $\pi_{C}(D)$, is the dataset obtained from $D$ by including only the columns $D_{\ast\f}$ of $D$ such that $\f$ is a feature of $S$ that satisfy $C$;  %\pmnote{how does a feature $\f$ satisfy a condition? probably need to say that $C$ is evaluated on all values of a $\f$ and if true, then $\f$ is included...}
    \item [$\sigma_C$:] the \emph{selection} of $D$ with respect to a boolean condition $C$ over $S$, denoted by $\sigma_C(D)$, is the dataset obtained from $D$ by including the rows $D_{i\ast}$ of $D$ satisfying $C$.
\end{description}
The condition of both the projection and the selection operators can refer to the values in $D$, as shown in the following example that uses an intuitive syntax for the condition.
\begin{example}
Consider the dataset $D$ in Example~\ref{ex:toyexample}. The result of the expression 
%$\pi_{\it \{features\ without\ nulls\}}(\sigma_{\ft{Gender}=F}(D))$ 
$\pi_{\it \{features\ without\ nulls\}}(\sigma_{\ft{Age}<30}(D))$ 
is the following dataset: 
\[%\small
\begin{array}{|c|ccc|}
\hline
\multicolumn{1}{|c|}{} & \ft{CId}& \ft{Gender} & \ft{Age}\\
\hline
    %\cline{2-5}
    1 & 113 & F & 24 \\ 
    %\cline{2-5}
    2 & 241 & M & 28\\
    \hline
\end{array}
\]
\end{example}

\myparagraph{Data augmentations.} Two basic data augmentation operators are defined over datasets. They allow the addition of columns and rows to a dataset, respectively.

\begin{description}
     \item [$\horaug_{f(X): Y}$:] the \textit{vertical augmentation} of $D$ to $Y$ using a function $f$ over a set $X=[\f_1 \ldots \f_k] \subseteq S$ of features, is obtained by adding to $D$ a new set of features $Y=[\f'_1 \ldots \f'_l]$ whose new values $d_{i\f'_1} \ldots d_{i\f'_l}$ for the $i$-th row are  obtained by applying $f$ to $d_{i\f_1} \ldots d_{i\f_k}$;  
     
     \item [$\veraug_{X:f(Y)}$:] the \textit{horizontal augmentation} of $D$ using an aggregative function $f$ 
     is obtained by adding one or more new rows to $D$ obtained by first grouping over the features in $X$ and then, for each group, by applying $f$ to $\pi_{Y}(D)$ (extending the result to $S$ with nulls if needed). 
     \added[id=pm]{Note that horizontal augmentation generates new rows based on grouping, i.e., by $X$, followed by an aggregation  $f(Y)$ applied to the values for $Y$ \textit{in each group}.}
\end{description}

\begin{example}\label{ex:augment}
Consider again the dataset $D$ in Example~\ref{ex:toyexample} and the following functions: (i) $f_1$, which associates the string \emph{young} when age is less than 25 and the string \emph{adult} otherwise, and (ii) $f_2$, which computes the average of a set of numbers.
Then, the expression
$\horaug_{f_1(\ft{Age}): \ft{ageRange}}(D)$ 
produces the following dataset: 
\[%\small
%\hspace{3mm}
\begin{array}{|c|cccc|c|}
\hline
\multicolumn{1}{|c|}{} & \ft{CId}& \ft{Gender} & \ft{Age}& \ft{Zip} & \ft{ageRange}\\
\hline
    %\cline{2-5}
    1 & 113 & F & 24 & 98567 & \textrm{young}\\ 
    %\cline{2-5}
    2 & 241 & M & 28 & \bot & \textrm{adult}\\ 
    %\cline{2-5}
    3 & 375 & C & \bot & 32768 & \bot \\
    %\cline{2-5}
    4 & 578 & F & 44 & 32768 & \textrm{adult}\\
    \hline
\end{array}
\]
\added[id=pm]{In expression
$E_2=\veraug_{\ft{Gender}:\mathit{avg}(\ft{Age})}(D)$ 
first \textit{group by} \ft{Gender} is computed, yielding two groups (for M and F), then $\mathit{avg}(\ft{Age})$ is executed on each group, resulting in the new rows 5,6 in the dataframe below:}

\[%\small
\begin{array}{|c|cccc|}
\hline
\multicolumn{1}{|c|}{} & \ft{CId}& \ft{Gender} & \ft{Age}& \ft{Zip}\\
\hline
    %\cline{2-5}
    1 & 113 & F & 24 & 98567\\ 
    %\cline{2-5}
    2 & 241 & M & 28 & \bot\\ 
    %\cline{2-5}
    3 & 375 & C & \bot & 32768\\
    %\cline{2-5}
    4 & 578 & F & 44 & 32768\\
    \hline
    5 & \bot & F & 34 & \bot\\
    6 & \bot & M & 28 & \bot\\
    \hline
\end{array}
\]
\end{example}
Note that new data can be added to a dataset using a horizontal augmentation where $X=\emptyset$, $Y=S$, and $f$ denote the procedure for adding records (e.g., by asking them to the user). Note also that horizontal augmentation allows us to combine, in the same dataset, entities at different levels of granularity, a feature that can be very useful to a data scientist (e.g., to compute, in the example above, the mean deviation).

\myparagraph{Data transformation.} One basic data transformation operator is defined over datasets:
\begin{description}
     \item [$\tau_{f(X)}$:] the \textit{transformation} of a set of features $X$ of $D$ using a function $f$ is obtained by substituting each value $d_{i\f}$ with $f(d_{\ast\f})$, for each feature $\f$ occurring in $X$.
%     \item [$\tau_{f:C}$:] the \textit{horizontal transformation} of $D$ using a function $f$ is obtained by substituting each value $d_{ij}$ in each row of $D$ satisfying $C$ with $f(d_{ij})$;
\end{description}

\begin{example} \label{ex:imputation}
Let $D$ be the dataset in Example~\ref{ex:toyexample} and $f$ be an imputation function that associates to the $\bot$'s occurring in a feature $\f$ the most frequent value occurring in $D_{\ast\f}$. Then, the result of the expression 
$\tau_{f(\ft{Zip})}(D)$ 
is the following dataset: 
\[%\small
\begin{array}{|c|cccc|}
\hline
\multicolumn{1}{|c|}{} & \ft{CId}& \ft{Gender} & \ft{Age}& \ft{Zip}\\
\hline
    %\cline{2-5}
    1 & 113 & F & 24 & 98567\\ 
    %\cline{2-5}
    2 & 241 & M & 28 & 32768\\ 
    %\cline{2-5}
    3 & 375 & C & \bot & 32768\\
    %\cline{2-5}
    4 & 578 & F & 44 & 32768\\
    \hline
\end{array}
\]
\end{example}

\myparagraph{Data fusion.} 
Given $D^L$ and $D^R$ on schemas $S^L$ and $S^R$ respectively, the two basic data fusion operators \textit{join} and \textit{append} allow the combination of a pair of datasets.

\begin{itemize}
     \item %[\it join:] 
     the \textit{join} $D^L \join^t_{C} D^R$ of $D^L$ and $D^R$ based on a boolean condition $C$ is the dataset over $S^L\cup S^R$ obtained by applying standard join operation of type $t$ (where $t$ can be equal to \textsf{inner}, (\textsf{left/right/full}) \textsf{outer}) based on the condition $C$;  
     
     \item %[\it append:] 
     the \textit{append} $D^L \union D^R$ of $D^L$ to $D^R$ is the dataset over $S^L\cup S^R$ obtained by appending  $D^L$ to $D^R$ and possibly extending the result with nulls on the mismatching columns $(S^L \cup S^R) \setminus (S^L \cap S^R)$.
\end{itemize}

\begin{example} \label{ex:fusion}
Let $D^L$ be the dataset in Example~\ref{ex:toyexample} (which we report here for convenience) and $D^R$ the dataset that follows. 
\[
%\small
D^L\!\!:
\begin{array}{|c|cccc|}
\hline
 & \ft{CId}& \ft{Gender} & \ft{Age}& \ft{Zip}\\
\hline
    %\cline{2-5}
    1 & 113 & F & 24 & 98567\\ 
    %\cline{2-5}
    2 & 241 & M & 28 & \bot\\ 
    %\cline{2-5}
    3 & 375 & C & \bot & 32768\\
    %\cline{2-5}
    4 & 578 & F & 44 & 32768\\
    \hline
\end{array}
\hspace{3mm}
D^R\!\!: 
\begin{array}{|c|cc|}
\hline
 & \ft{CId}& \ft{name}\\
\hline
    %\cline{2-5}
    1 & 241 & \textrm{Jim}\\ 
    %\cline{2-5}
    2 & 578 & \textrm{Mary}\\
    \hline
\end{array}
\]

Then, the result of the expression 
$D^L\join^{\sf inner}_{D^L.\ft{CId}=D^R.\ft{CId}} D^R$ 
is the following dataset: 
\[%\small
\begin{array}{|c|ccccc|}
\hline
\multicolumn{1}{|c|}{} & \ft{CId}& \ft{Gender} & \ft{Age}& \ft{Zip}& \ft{Name}\\
\hline
    1 & 241 & M & 28 & \bot & \textrm{Jim}\\ 
    2 & 578 & F & 44 & 32768 & \textrm{Mary}\\
    \hline
\end{array}
\]
On the other hand, the result of the expression 
$D^L\union D^R$ 
is the following dataset: 
\[%\small
\begin{array}{|c|ccccc|}
\hline
\multicolumn{1}{|c|}{} & \ft{CId}& \ft{Gender} & \ft{Age}& \ft{Zip}& \ft{Name}\\
\hline
    %\cline{2-5}
    1 & 113 & F & 24 & 98567 & \bot\\ 
    2 & 241 & M & 28 & \bot & \bot\\ 
    3 & 375 & C & \bot & 32768 & \bot\\
    4 & 578 & F & 44 & 32768 & \bot\\
    5 & 241 & \bot & \bot & \bot & \textrm{Jim}\\ 
    6 & 578 & \bot & \bot & \bot & \textrm{Mary}\\ 
    \hline
\end{array}
\]
\end{example}

We note that the data manipulation model presented here has some similarities with the Dataframe algebra~\cite{PetersohnMLMXMG20}. The main difference is that we have focused on a restricted set of core operators (with some of that in~\cite {PetersohnMLMXMG20} missing and others combined in one) with the specific goal of providing a solid basis to an effective technique for capturing data provenance of classical preprocessing operators. We point out that our algebra can be easily extended to include operators implementing other ETL/ELT-like transformations whose fine-grained provenance capture has been described elsewhere \cite{article}. \label{dataframe}

%In the following, we will call \emph{DM expression} (where DM stands for data manipulation) a combination of the above operators and by $E(D)$ the application of a DM expression $E$ to a dataset $D$.

\subsection{Data provenance model}
\label{sec:provenance-model}

\begin{figure}[ht]
\centering
\includegraphics[width=.4\textwidth]{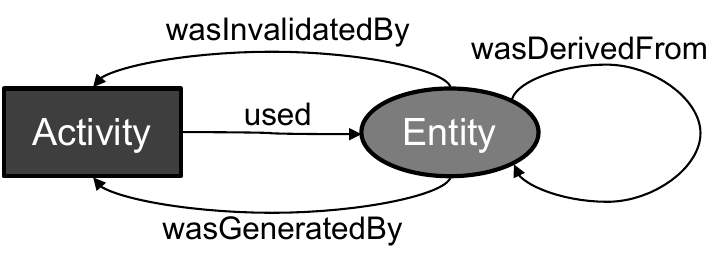}
\caption{The core W3C PROV model.}
\label{fig:provmodel}
\end{figure}

The purpose of data provenance, in this setting, is to support the generation of  simple explanations for the existence (or the absence) of some piece of data in the result of complex data manipulations. 
Along this line, we adopt as the provenance model a subset of the PROV model~\cite{provw3c} from the W3C, a widely adopted ontology that formalises the notion of \textit{provenance document} and which admits RDF and other serialisation formats to facilitate interoperability. The minimal elements of the model are graphically described  as shown in Figure~\ref{fig:provmodel}.

In PROV an entity represents an element $d$ of a dataset $D$ and is uniquely identified by $D$ and the coordinates of $d$ in $D$ (i.e., the corresponding row index and feature). An activity represents any pre-processing data manipulation that operates over datasets. 
For each element $d$ in a dataset $D'$ generated by an operation $\bf o$ over a dataset $D$ we represent the facts that:  (i) $d$  \textit{wasGeneratedBy} $\bf o$, and (ii)  $d$  \textit{wasDerivedFrom} a set of elements in $D$. In addition, we represent: (iii) all the elements $d$ of $D$ such that $d$ was \textit{used} by $\bf o$ and (iv) all the elements $d$ of $D$ such that $d$  \textit{wasInvalidatedBy} (i.e., deleted by) $\bf o$ (if any).
Note that in PROV derivation implies usage, but the inverse is not true and this is why this notation is not redundant.

%\[D_{1,\ft{Age}}:24\]
%\[D_{2,\ft{Age}}:28\]
%\[D'_{1,\ft{ageRange}}:young\]
%\[D'_{2,\ft{ageRange}}:adult\]
%\[E=\horaug_{f_1:\ft{Age}\mapsto \ft{ageRange}}\]

\begin{example} \label{ex:ProvModel}
Let $E$ be the first expression in Example~\ref{ex:augment} and $D'=E(D)$. A fragment of the data provenance generated by this operation, for two of the dataset elements, is reported in Figure \ref{fig:exProvModel}.

\begin{figure}[ht]
\centering
\includegraphics[width=.6\textwidth]{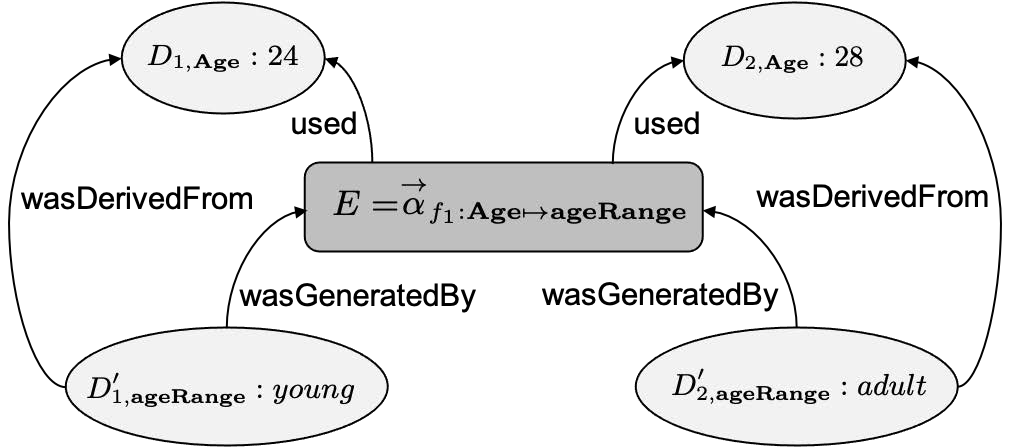}
\caption{A fragment of provenance data for the operation in Example~\ref{ex:ProvModel}.}
\label{fig:exProvModel}
\end{figure}
\end{example}

\subsection{\added[id=rt]{Limitations and possible extensions}}
\label{sec:limitation}
\added[id=rt]{
The models for data representation, manipulation, and provenance generation introduced in the previous sections cover a large body of data preparation pipelines, but they are clearly not exhaustive. In particular, we have assumed that the input data is in a bi-dimensional, tabular format (e.g., csv files), with rows representing observations of some phenomena and columns representing interesting features of the observations. However, multi-dimensional data,  including tensors and matrices, are common in many machine learning applications. 
Our model can be extended by assuming that each value is indeed a measure for a combination of features, possibly at different levels of aggregation, similar to logical multidimensional data models that have been proposed in the literature for data warehousing and OLAP (e.g.,~\cite{CabibboT98}). 
This would also make it possible to include multi-level aggregation operations (roll-up, drill-down, slicing, and dicing) by extending the data model as described e.g. in ~\cite{CabibboT98}.
These extensions add complexity to the resulting provenance graphs, as the derivations must be traced through multiple aggregations, however, this would not add to our conceptual framework, as these simply extend the fundamental provenance patterns used for standard one-level aggregations (see Sec.\ref{sec:ha}). Supporting such extensions is therefore currently beyond the scope of our proof-of-concept implementation.}
% \item \added[id=rt]{We have considered only a subset of the PROV model~\cite{provw3c} for representing provenance data. Further features of this model could be easily added to the framework but at the cost of an increasing volume of provenance data generated for each operation. We therefore focus on the components of provenance that are strictly needed for our purposes, as described in the following section.

% \end{itemize}

\subsection{Problem Statement} 
\label{sec:statement}

We consider compositions of the operators introduced in Section~\ref{sec:dataman} into \textit{pipelines} that 
take as input a collection of datasets $D_1,\ldots,D_n$ and produce a dataset $D'$, denoted $D' = p(D_1,\ldots,D_n)$, by applying a (partially ordered) sequence of operators. Note that $E$ can be represented as a tree, where the internal nodes are operators and leaves are datasets.
Note also that although in principle any combination is possible, in practice there are constraints on the ordering of the operators, because some operators may alter the dataset schema.

The performance of the  model learned from  $p(D)$ is dependent upon the 
operators involved in $p$. As the data scientist iterates over versions of the models, they may wish to inspect and understand exactly what happened at each step within the pipeline. This can be a complex manual task for any realistic pipeline.  

The provenance collected by the system presented here is intended to allow the data scientist to review, understand, and debug what happened in past runs of any given pipeline. Depending on the granularity of the provenance, this can be as coarse-grained as a dataset, $p$ was used with transformations $T_1, T_2, ...$ to a very fine-grained version which allows users to track individual data items as described previously. 

Classic provenance queries include: Why~\cite{buneman_why_2001}, How~\cite{buneman_how_2009} and Why Not~\cite{chapman2009}. Instances of each of these queries are shown in Table \ref{tab:provqueries} as Queries 2, 3, and 7-9 respectively.  In addition to these classic provenance queries, we have analyzed questions posed to the Data Science Stack Exchange (DSSE) about problems posed by users, encountered when trying to understand and debug the pipelines. An explanation of the use cases and the provenance queries in Table \ref{tab:provqueries} that they relate to can be found in Table~\ref{tab:realprovqueries}. Through this analysis, we have identified an additional 6 provenance queries based on the use cases from DSSE: All Transformations (1); Dataset-level Feature Operation (4); Record Operation (5); Item-level Feature Operation (6); Impact on Feature Spread (12) and Impact on Dataset Spread (13). Queries 1, 4, and 5 are similar to How-provenance but are focused only on the transformations. The difference between them is the granularity of focus - dataset, feature, record, or individual value. Queries 10 and 11 have been implemented to emphasize how provenance can help when developing a pipeline. They show what an item was and what it will be, highlighting potential errors or imperfections. Queries 12 and 13, however, present a new usage of provenance, and thus a distinctly new provenance query type. 

In the DSSE use cases, it became clear that a question being asked was ``what operations were performed to the data and how did those change the data profile''. This is a reasonable question as these transformations may entail unintended consequences, as they alter the representation of the domain that  the learning algorithms generalize from, and they may remove or inadvertently introduce new bias in the data \cite{ghorbani2019data}. 
In turn, this may reflect on non-performance properties of the models, such as their \textit{fairness}. Fairness, formally defined in terms of statistical properties of the model's predictions~\cite{narayanan2018translation}, broadly refers to the capability of a model to ensure that its predictions are not affected by an individual belonging to one of the groups defined by some sensitive attribute(s), such as sex, ethnicity, income band, etc. Queries 12 and 13 provide a mechanism that computes the statistical properties of the data before and after an operation to identify when there are major shifts in distributions.
 
Thus, the problem within this work is to: a) define the set of operations for data manipulation available within a pipeline; b) establish a set of provenance templates that can be used to reason over and capture the provenance of these operations over the data; c) show that our approach can support typical provenance queries in an effective and scalable way.

\begin{table*}[t]\footnotesize
\caption{Provenance queries of interest to a data scientist designing a pipeline of pre-processing operations.}
%\vspace*{-3mm}
\label{tab:provqueries}
\centering%\small
    \begin{tabular}{|c|c|c|p{7.5cm}|}
         \hline
        \textbf{Id} & \textbf{Provenance Query}  & \textbf{Input} & \textbf{Output}  \\
         \hline\hline
        PQ1 &  \textrm{All Transformations}  & $D$ & Set of operations applied to $D$ and the features they affect. \\
         \hline
        PQ2 & \textrm{Why-provenance} & $d_{i\ft{a}}$ & The input data that influenced $d_{i\ft{a}}$.  \\
         \hline
        PQ3 & \textrm{How-provenance}  & $d_{i\ft{a}}$ & The input data and the operations that created $d_{i\ft{a}}$. \\
         \hline
        PQ4 & \textrm{Dataset-level Feature Operation} &  $D_{\ast\ft{a}}$ & Set of operations
        %that were applied to the feature $D_{\ast\ft{a}}$.  & \multirow{3}{*}{UC6}\\
        that were applied to feature $\ft{a}$. \\
         \hline
        PQ5 & \textrm{Record Operation}  & $D_{i\ast}$ & Set of operations that were applied to record $D_{i\ast}$.  \\
         \hline
        PQ6 & \textrm{Item-level Feature Operation}  & $d_{i\ft{a}}$ & Set of operations that were applied to $d_{i\ft{a}}$. \\
         \hline
        PQ7 & \textrm{Feature Invalidation} & $D,\ft{a}$ & The operation that deleted the feature $\ft{a}$. \\
         \hline
        PQ8 & \textrm{Record Invalidation}  & $D,i$ & The operation that deleted the record $D_{i\ast}$. \\
         \hline
        PQ9 & \textrm{Item Invalidation}  & $D,i,\ft{a}$ & The operation that deleted the item $d_{i\ft{a}}$.  \\
         \hline
        PQ10 & \textrm{Item History}  & $d_{i\ft{a}}$ & All the elements derived and that will derive from $d_{i\ft{a}}$. \\
         \hline
        PQ11 & \textrm{Record History} & $D_{i\ast}$ & All the  elements derived and that will derive from $D_{i\ast}$.  \\
         \hline
        PQ12 & \textrm{Impact on Feature Spread} & $D$ & The change in feature spread of all  operations over a feature of $D$. \\
         \hline
        PQ13 & \textrm{Impact on Dataset Spread} & $D$ & The change in dataset spread of all operations applied to $D$. \\
         \hline
    \end{tabular}
\end{table*}{}

\section{Data processing operators} \label{sec:operators}

In this section, we illustrate a number of common data pprocessing operators that are often used in data preparation workflows, showing how they can be suitably expressed as a composition of the basic operators introduced in Section~\ref{sec:dataman}.

\subsection{Data Reduction}

\myparagraph{Feature Selection.} This operation consists of selecting a set of relevant features from a given dataset and dropping the others, which are either redundant or irrelevant to the goal of the learning process.

Feature selection over a dataset $D$ with a schema $S$ can be expressed by means of a simple pipeline involving only the projection operator with a condition that selects the set of features $I\subset S$ of interest:
\[
\it FS(D)=\pi_C(D)
\]
where $C=\{\it \f \in I\}$.

A special case of feature selection is an operation that drops columns with 
a value rate of missing values higher than a threshold $t$. In this case, the condition of the projection operator is more involved as it requires introspection of the dataset:
\[
C=\{\it \f \in S \mid \textsf{count}(D_{i\f}=\bot, 1\leq i\leq n) < t \}.
\]
    
\myparagraph{Instance Selection.} The aim of this operation is to reduce the original dataset to a manageable volume by removing certain records with the goal of improving the accuracy (and efficiency) of classification problems.

Also in this case, instance selection over a dataset $D$ with a schema $S$ can be expressed by means of a simple pipeline involving only the selection operator with a condition that identifies the set of relevant rows of $D$ by means of a predicate $p$:
%\[
$\it IS(D)=\sigma_C(D)$
%\]
where $C=\{\it D_{i\ast} \in S \mid p(D_{i\ast})\}.$

Similar to feature selection, a relevant case of instance selection drops rows with 
a value rate of missing values higher than a threshold $t$. In this case, 
\[C=\{\it D_{i\ast} \in D \mid \textsf{count}(D_{ij}=\bot, 1\leq j\leq m) < t \}\]

\subsection{Data Transformations}
By data transformation, we mean any operation on a given dataset that modifies its values  with the goal of improving the quality of $D$ and/or making the process of information extraction from $D$ more effective. 
The following operator is meant to capture data transformation ($\mathit{DT}$ in its generality:
\[
\mathit{DT}(D)=\tau_{f(X)}(D)
\]
where $f$ is any scalar function that generates a new value $f(x)$ from values of feature set $X$ of $S$.
Several cases of transformations are common in pre-processing pipelines, as illustrated in the following.
         
\myparagraph{Data repair.}
It is the process of replacing inconsistent data items with new values. In this case, $f$ is a simple function that converts values and the data transformation possibly operates on the whole dataset. 

\myparagraph{Binarization.} It is the process of converting numerical features to binary features. For instance, if a value for a given feature is greater than a threshold it is changed a 1, if not to 0.

\myparagraph{Normalization.} It is a scaling technique that transforms all the values of a feature so that they fall in a smaller range, such as from 0 to 1. There are many normalization techniques, such as Min-Max normalization, Z-score normalization, and Decimal scaling normalization. This operation operates on a single feature at a time

\myparagraph{Discretization.} It consists of converting or partitioning continuous features into discrete or nominal features. It performs a value transformation from categorical to numerical data.

\myparagraph{Imputation.} It is the process of replacing missing data (nulls in our data model) with valid data using a variety of  statistical approaches that aim at identifying the values with the maximum likelihood.

\subsection{Data augmentation}
\label{sec:da}

\myparagraph{Space Transformation.}
This operation takes a set of features of an existing dataset and generates from them a new set of features by combining the corresponding values. Usually, the goal is to represent (a subset of) the original set of features in terms of others in order to increase the quality of learning.

The application of this operation to a dataset $D$ over a schema $S$ can be expressed by means of an expression involving a vertical augmentation that operates on a subset $X$ of the features in $S$ and produces a new set of features $Y$, followed by a projection operator that eliminates the features in $X$, 
thus maintaining those in $Z=(S\cup Y)-X$: 
\[
\it 
ST(D)=\pi_{\{\it features\ in\ Z\}}(\horaug_{f(X): Y}(D))\label{ch:st}
\]

\myparagraph{Instance Generation:}
\added[id=pm]{These operators include grouping and aggregation, and their effect is to fill regions in the domain of the problem, which does not have representative examples in original data, or to summarize large amounts of instances in fewer examples.}
These are also denoted \textit{prototype generation} methods, as the artificial examples created tend to act as a representative of a region or of a  subset of the original instances.

The application of this operation to a dataset $D$ over a schema $S$ can be expressed by means of an expression involving a horizontal augmentation 
that, if needed, groups over a subset $X$ of the features in $S$ and 
then apply a summary function $f$ over another subset of $S$:
\[
\it IG(D)=\veraug_{X:f(Y)}(D).
\]

This operation can be preceded by a data reduction operator (a projection or a selection) to isolate the portion of the original dataset on which we intend to operate.

\myparagraph{String Indexer.} This operator encodes a feature involving strings into a feature of string indices. The indices are in $\it [0, numLabels)$.  It is a special case of Space transformation.

\myparagraph{One-Hot Encoder.} 
This operation maps a feature involving strings to a set of boolean features.  Specifically, it creates one column for each possible value occurring in the feature. Each new feature gets a 1 if the row contained that value and a 0 if not. It is a special case of space transformation.

\subsection{Data fusion}

Data preparation pipelines often require combining datasets coming from different data sources. For this reason, packages for data pre-processing are usually equipped with facilities for combining datasets in two main ways, as follows.

\myparagraph{Data integration.} It is the process of combining rows of two datasets on the basis of some common property. This can be useful when, for instance, we need to extend the features of observations of a phenomenon or objects of interest \added[id=rt]{stored in a dataset $D_1$} (e.g., the technical information of smartphones on sale) with further features of the same observations or of the same objects gathered elsewhere \added[id=rt]{stored in a dataset $D_2$} (e.g., the ratings of the same smartphones available on a review site). 
\added[id=rt]{This activity can be supported by an expression involving the join operator over the datasets under consideration and can be preceded by a data reduction operator to isolate the portion $X$ of the original dataset on which we intend to operate, as follows:
\[\pi_{X}(D_1 \join^{\it left}_{C} D_2)\]
where $C$ specifies the condition that rows of different datasets must satisfy to be combined (e.g., they share the same standardized product identifier).}
\deleted[id=rt]{The join operator $\join^t_{C}$, in which $C$ specifies the condition that rows of different datasets must satisfy to be combined (e.g., they share the same standardized product identifier), serves for such purposes. 
This operation can be preceded by a data reduction operator (a projection or a selection) to isolate the portion of the original dataset on which we intend to operate.}
The join operator can also be equipped with some sophisticated techniques for joining rows, such as one based on entity resolution. 

\myparagraph{Data expansion.} It is the process of putting together rows of two datasets that contain data referring to different observations of the same phenomenon or to different objects of the same type. This can be useful when, for instance, a training set is built by accumulating data coming from diverse data sources, \added[id=rt]{say $D_1$, $D_2$ and $D_3$} (e.g., experimental data of a medical treatment produced by \added[id=rt]{three} different laboratories). The append operator $\union$ can be used in such scenarios \added[id=rt]{possibly preceded by some data reduction operators, for example as follows:
\[D_1 \union \pi_{C_1}(D_2) \union \sigma_{C_2}(D_3)\]}. 
As shown in Example \ref{ex:fusion}, this operator also accounts for situations in which we need to merge datasets that involve different features of the same phenomena.

\section{Abstract analysis of provenance capture} \label{sec:provenance}

In order to capture the provenance of a pipeline $p$ of a combination 
of pre-processing operations $\oo_1,\ldots,\oo_n$ forming a tree,  we introduce an abstract provenance-generating function (\textit{prov-gen}), and associate it with each operation $\oo_k$ occurring in $p$. 

In accordance with the provenance model presented in Section \ref{sec:provenance-model},  each element $d_{ij}$ of a dataset $D$ produced during the execution of  $p$ is represented by a PROV entity in the provenance document. The properties of this entity include the row index $i$ and feature $j$ in $D$, and an identifier $k$ denoting the fact that $d_{ij}$
is in the result of the operation $\oo_k$ in $p$.

Similarly, each operation $\oo_k$ in $p$ is represented by a PROV \textit{activity} in the provenance document, whose properties specify the operator(s) illustrated in Section \ref{sec:dataman} that implement(s) $\oo_k$, and the list of the features on which $\oo_k$ operates.

\subsection{Provenance templates}
\label{sec:templates}

We now present example instances of  provenance-generating (\textit{prov-gen}) functions  for the main types of operations observed in data science pipelines, discussed in Section \ref{sec:operators}.  
To recall, these are: (i) data reduction:
$D' = \pi_C(D)$,  $D' = \sigma_C(D)$;
(ii) Data augmentations:  $\horaug_{f(X): Y}$, $\veraug_{X:f(Y)}$;
 (iii) Data transformations: $\tau_{f(X)}$; and 
 (iv) Data fusion.

A prov-gen function takes as inputs the sets of input and output values $D,D'$ for the operator, and produces a PROV document that describes the transformation produced by the operator on each element of $D$, as reflected in $D'$. Note that for binary operators, namely join and append, $D$ includes inputs from both operands.

Take for example the case of Vertical Augmentation (VA):
$\horaug_{f_1(\ft{Age}): \ft{ageRange}}(D)$ 
which we used in Example~\ref{ex:augment}, where attribute \textbf{Age} is binarised into \textit{\{young, adult\}} based on a pre-defined cutoff, defined as part of $f()$.
The prov-gen function for VA will have to produce a collection of small PROV documents, one for each input-output pair $\langle D_{i,\ft{Age}}, D'_{i,\ft{AgeRange}} \rangle$ as shown in the example.

As these documents all share the same structure, we define a common  \textit{PROV template} which is then instantiated multiple times, once for each input/output pair.
A template is simply a PROV document that may contain variables, indicated by the namespace \textbf{var:}, which are used as placeholders for values. Here templates are designed to capture the transformation at the level of individual elements of $D$, or its rows or columns, as appropriate. Thus a template will have a \emph{used} set of entities, which refer to the subset of data items in $D$ which have been used by $\bf o$, and a \emph{generated} set of new entities, corresponding to new elements in $D'$ (for projection and selection, it will have an \emph{invalidated} set of entities instead, as these operators remove data from $D$).

\begin{figure}[ht]
    \includegraphics[width=.7\columnwidth]{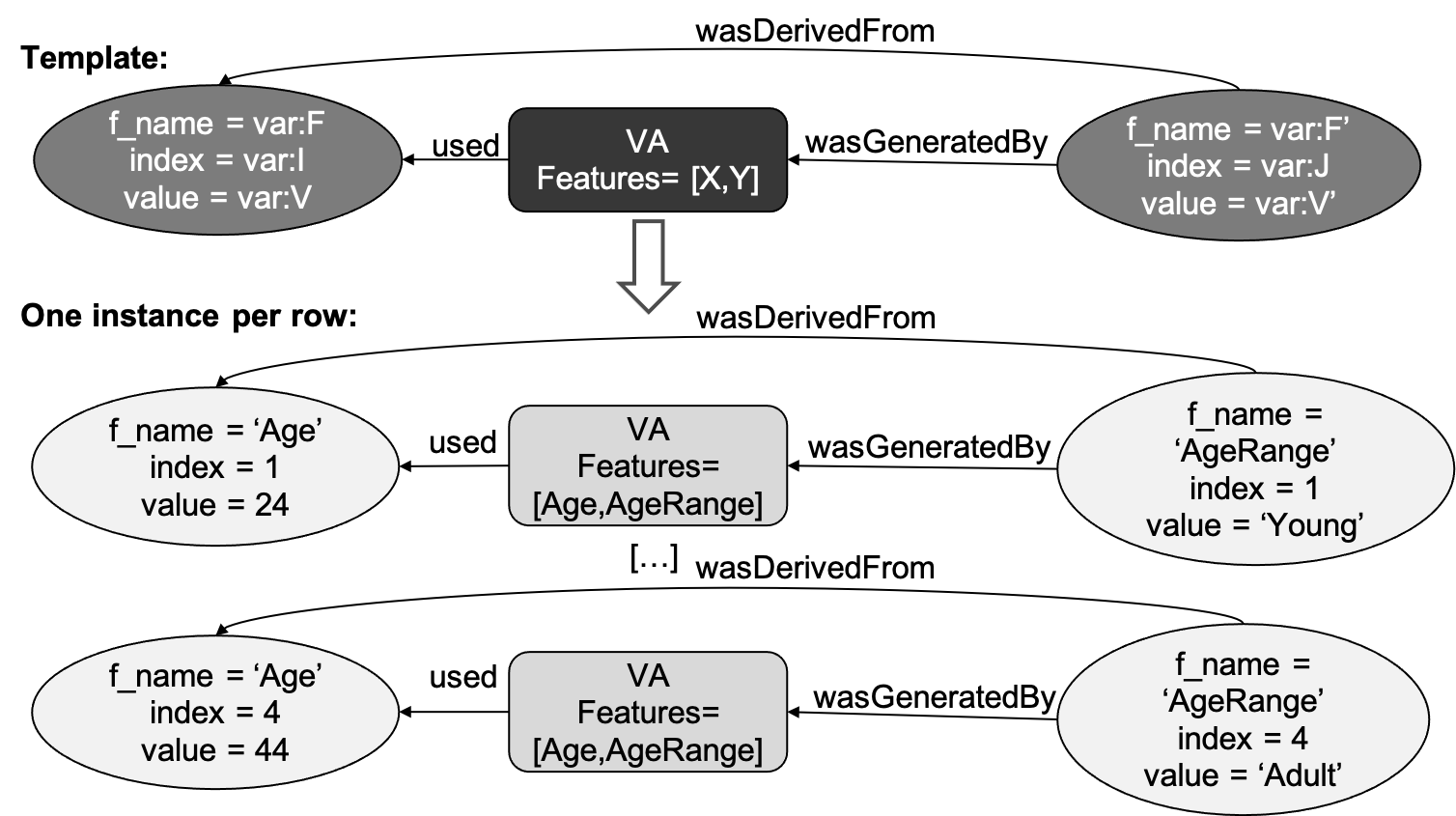}
    \centering
\caption{Example of PROV template for Vertical Augmentation and corresponding instances.}
\label{fig:template-example}\end{figure}

The PROV template for (VA) is shown in Figure~\ref{fig:template-example}, where we use the generic attribute names $X,Y$ to indicate the old and new feature names.
One or more \textit{binding generators} are associated with each template: they determine how values found in $D$, $D'$ upon execution of the operator are substituted for the variables. Each variable substitution results in a \added[id=pm]{small PROV document, which represents all derivations through a single operator.}
\replaced[id=pm]{In the following we are going to refer informally to these documents as \textit{provlets}, to indicate that a complete PROV document representing a complex  derivation chain can be produced by joining multiple such provlets on their data identifiers, as described in Sec~\ref{sec:generation} below.}{which we refer to as a \textit{provlet}}

In the VA example, the transformation between $D$ and $D'$ is 1:1 and thus a new provlet is created from  each value $D_{*,\ft{Age}}$ of column $\ft{Age}$ and the corresponding value in $ \ft{AgeRange}$.

Using a list comprehension notation, the binding generator for the variables used in the template in Figure \ref{fig:template-example}\label{vartemplate} are defined as:
\begin{align*} \small
[ \langle & F = \ft{Age}, I=i, V=D_{i,\ft{Age}}, %\\
          %& 
          F' = \ft{AgeRange} , J=i, V'=f(D_{i,\ft{Age}}) \rangle |  i:1 \dots n ]       
\end{align*}
These are the new entities for the newly created data elements in the new column $D_{*,\ft{AgeRange}} \in \{\textit{young}, \textit{adult}\}$.
Two of the $n$ PROV documents for this  specific example are shown in Figure~\ref{fig:template-example}.

\subsection{Template binding rules}\label{sec:binding}

\begin{figure}
    \includegraphics[width=.7\columnwidth]{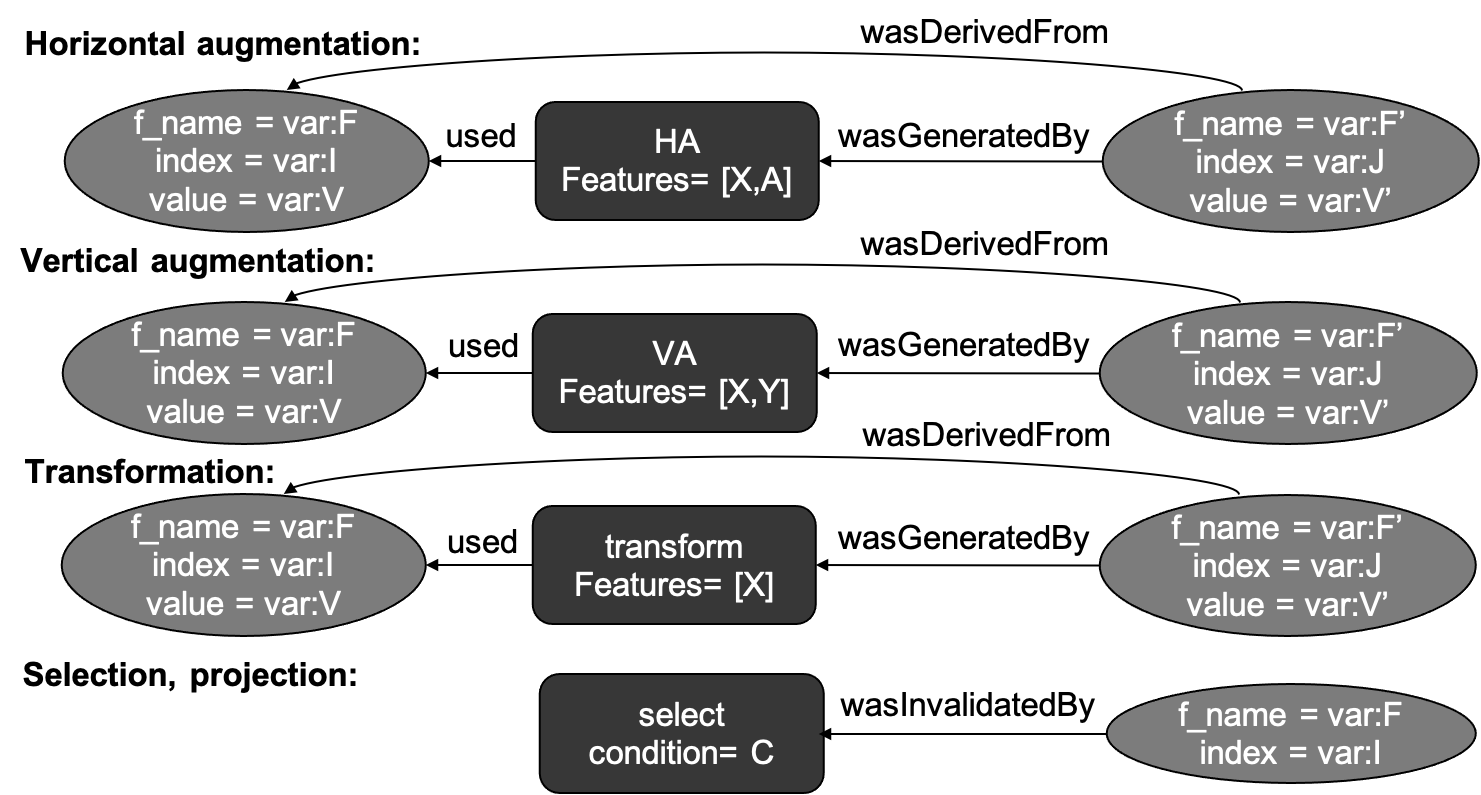}
    \centering
\caption{PROV templates used by the prov-gen functions for data augmentation, transformation, and reduction.}
\label{fig:all-templates}\end{figure}

We define templates for each of the five core operators, shown in Figure~\ref{fig:all-templates} and the corresponding binding generators for \textit{used}, \textit{generated}, and \textit{invalidated} sets of entities.

Note that we do not need to create complete provlets for all entities in any given output dataset. If $f(D)$ does not change $d_{ij}$,  then no provenance record needs to be generated. However if $f(D)$ discards elements of $D$, then a provlet containing an invalidation relationship is required. Whenever a new entity is generated, i.e. when $f(D)$ creates a  new or updated value in $d_{ij}$, a complete provlet is also required. In other words, we only require provenance statements that capture different versions  between elements in the dataset.

\subsubsection{Data reduction, selection} \label{sec:data-reduction-selection}
Data reduction \textit{invalidates} existing entities. For selection: $D' = \sigma_C(D)$, the bindings specify that an entire row $i$ is invalidated whenever condition $C$ is False when evaluated on that row. This affects all features $X \in S$:
$$ [ \langle F=X, I=i\rangle | X \in S, i:1\dots n, C(D_{i,*} = \text{False}) ] $$
A \textit{wasInvalidatedBy} relationship is established between each of these entities and a single \textit{Activity}, representing the selection.

\subsubsection{Data reduction, projection} Conditional projection $D' = \pi_C(D)$ invalidates all elements in column $X \in S$   whenever $C$ returns True when evaluated on elements of $X$:
\begin{align*}
& [ \langle F=X, I=i\rangle | X \in S , i:1\dots n, C(D_{*,X} = True )]
\end{align*}
Similar to the selection, here too a \textit{wasInvalidatedBy} relationship is established between each of these entities and a single \textit{Activity}, representing the projection.

\subsubsection{Vertical augmentation} \label{sec:va}
$\horaug_{f(X): Y)}$ takes a set $X \subset S$ of features and adds a new set  $Y$ of features, $Y \cap S = \emptyset$ to $D'$ as shown in Ex.~\ref{ex:augment}. The provenance consists of $n$ PROV documents, one for each row $i$ of $D$, and in each such document entities for $D_{i,X_m}, X_m \in X$ are used to generate entities for the new features $Y_h \in Y$. Thus, the bindings are defined as follows:
\begin{align*}
& \text{For } i:1 \dots n:&\\
&\ \ \ \text{\textit{used} entity:} [\langle F = X_m, I=i, V=D_{i,X_m} \rangle |  X_m \in X ] \\
&\ \ \ \text{\textit{generated} entity:}  [\langle F' = Y_h, J=i, v=f(D_{i,X}) \rangle |  Y_h \in Y ]
\end{align*}
These entities are then connected to a single \textit{Activity}, as shown in Figure~\ref{fig:all-templates} and in the examples (Figg.~\ref{fig:template-example}, ~\ref{fig:example-composition}), using \textit{Used} and \textit{wasGeneratedBy} relationship.
For each pair of \textit{used}, \textit{generated} entities having the same index on each side (i.e., where \textbf{var:I} = \textbf{var:J} after template instantiation), a \textit{wasDerivedFrom} relationship is also added, to assert a stronger relationship (derivation occurs through the Activity that connects the entities).

\subsubsection{Horizontal augmentation, \added[id=pm]{grouping and aggregation}} \label{sec:ha}
The $\veraug_{X:f(Y)}$ operator groups records according to columns $X \subset S$, producing a list $G = [g_1 \dots g_h]$ of $h$ groups. Then for each  $g_i \in G$ it computes  $f(Y)$ from the records in the group, producing a new record containing the aggregated value in column $A$, the values that define the group in each column $X_m \in X$, which we denote $\mathit{val(X_m, g_i)}$, and \textit{null} in all other columns  (see Ex. ~\ref{ex:augment} in Section~\ref{sec:dataman}).
Thus, the operator produces $h$ records, and let $\mathit{rows}(G) = [n+ 1,n+2,...n+h]$ denote their new row indexes in the dataset.

\added[id=pm]{The corresponding provenance template and binding rules are similar to those for Vertical Augmentation (Figure ~\ref{fig:all-templates}), but with some differences, and are best illustrated initially using an example. Consider the following dataframe:}

\[%\small
\begin{array}{|c|ccc|}
\hline
%\multicolumn{1}{|c|}{} & \ft{X_1} & \ft{A} & \ft{B}\\
 &\bf X_1 &\bf A &\bf B\\
\hline
    1 & x_1  & 10 & b_1 \\ 
    2 & x_2 & 30 & b_2 \\ 
    3 & x_1 & 20 & b_3 \\
    4 & x_2 & 40 & b_4 \\
    \hline
    5 & x_1 & 30 & \bot \\
    6 & x_2 & 70 & \bot \\
    \hline
\end{array}
\]

\added[id=pm]{and grouping operator $\veraug_{X:f(Y)}$ where $X = [X_1]$, $f(Y)= \sum{A}$, where the sum on $A$ values occurs for each group.}

\begin{figure}
    \includegraphics[width=.85\columnwidth]{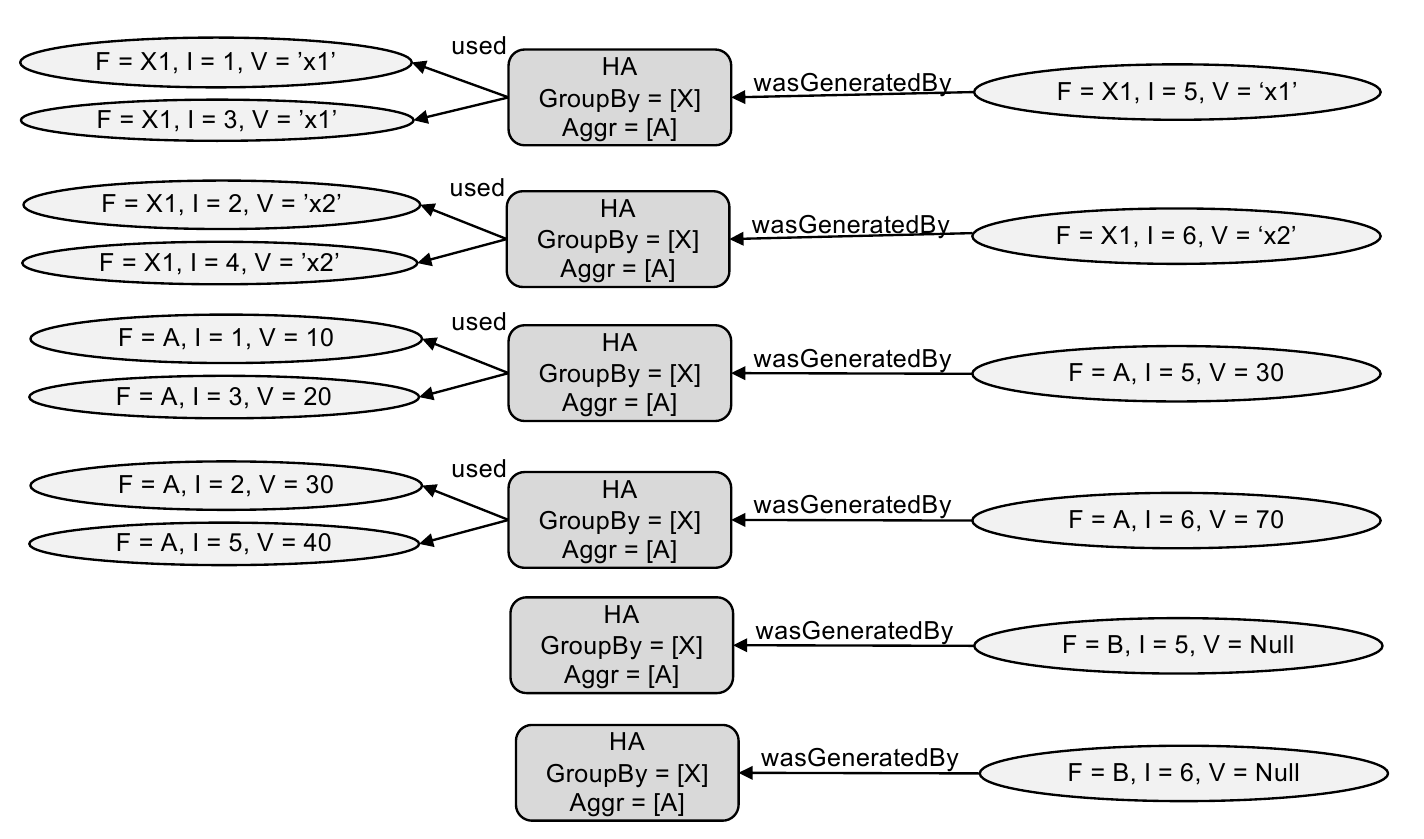}
    \centering
\caption{Provlets describing grouping and aggregation from Example in Sec.\ref{sec:ha}}
\label{fig:HA}\end{figure}

\added[id=pm]{The set of provlets that represent the derivations of the elements in the new rows 5, 6 are depicted in Fig.~\ref{fig:HA}.
Values $x_1, x_2$ in rows 5,6 identify the groups and are derived from the corresponding values in rows 1,3 and 2,4, respectively. 
Similarly, the two values in column A are obtained by adding up  the corresponding A values in the same groups of rows (1,3 and 2,4).
Finally, the null values in column B are generated by the operator, but their values are not derived from any inputs.}

\added[id=pm]{Generating these provlets requires maintaining the association between each group: group 1 in row 5, and group 2 in row 6, and the corresponding input rows (1,3 and 2,4). 
Implementations can achieve this in different ways. Formally, we assume that each group $g_i \in G$ maps to a set of ``group input'' rows, i.e., $\mathit{ginput}(g_i)$. In our example, we have 
$\mathit{ginputs}(g_1) = \{1,3\}, \mathit{ginput}(g_2) = \{2,4\}$.
}

\added[id=pm]{Then, the binding rule for group $g_i$ in row $i$ and elements in each of the columns $C \in X$ can be written as follows.}
\added[id=pm]{For generated entities:
\[ \langle F = C, I=i, V=D_{i,C} \rangle \]
For used entities:
\[ [\langle F = C, I=j, V=D_{i,C} \rangle |  j \in \mathit{ginput}(g_i) ] \]}

\added[id=pm]{Similarly, for the values in columns $C' \in Y$, the rule for generated entities is:
\[ \langle F = C', I=i, V=D_{i,C'} \rangle \]
and for used entities:
\[ [\langle F = C', I=j, V=D_{i,C'} \rangle |  j \in \mathit{ginput}(g_i) ] \]
Finally, the rule for the null values applies to values in columns $C" = S \setminus Y \setminus X$, and they only have the \textit{generatedBy} side:
\[ \langle F = C, I=i, V=Null \rangle \]
}

\begin{figure}
    \includegraphics[width=.85\columnwidth]{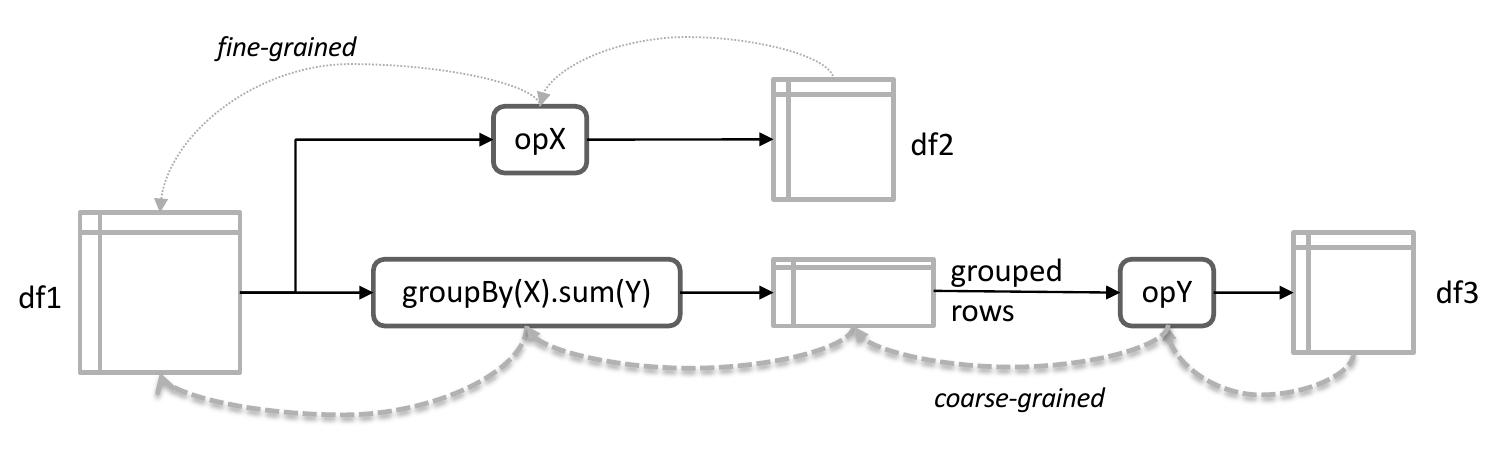}
    \centering
\caption{Aggregation pattern and loss of provenance granularity}
\label{fig:grouping-granularity}\end{figure}

\added[id=pm]{Note that aggregation operators reduce the granularity of the derivations. Typical Value Transformation operators, for instance a normaliser, would map each input element to a corresponding output element. Aggregations, on the other hand, produce a ``provenance bottleneck'' where $n$ rows are mapped to $m<n$ rows, where $m$ is the number of groups, because the provenance of any ``downstream'' dataframe that makes use of the groups will have to include one of the group rows.
In practice, aggregations may produce a pipeline pattern as shown in Fig.~\ref{fig:grouping-granularity}, where some of the operators (\texttt{opY}) use the aggregations, and the provenance of new dataframe elements produced by these operators will map to grouped rows and not to the upstream un-aggregated dataframes, leading to some loss of granularity.
In the Figure, the bottom provenance dependencies (thick dotted lines) for elements of \texttt{df3} must include some of the group rows, and those in turn are derived from \textit{each} of the inputs. 
Note also, however, that the loss of granularity depends on the number of groups. In the extreme case where the grouping operator produces a single group consisting of all input rows, for instance, the result is a provenance graph where all inputs contribute to the grouping, and all outputs depend on the grouping, producing a complete bottleneck.} 

% \textit{Used} entities.
% For each  $g_i\in G$, let $\mathit{rows}(g_i)$ denote the set of row indexes for records in $g_i$. The bindings associated with $g_i$ are:
% \[ [\langle F = A, I=i, V=D_{i,A} \rangle |  i \in \mathit{rows}(g_i) ]. \]

% \textit{Generated} entities. 
% For each $g_i$, the new record with index $n+i$ is represented by a set of generated entities, with bindings:
% \begin{align*}
% &[\langle F' = A, I=i, V=f(D_{rows(g), A}) \rangle ] \\
% &[\langle F' = Y, I=i, V=\mathit{null} \rangle | Y \in S \setminus X, Y\neq A] \\
% &[\langle F' = X_m, I=i, V=\mathit{val(X_m, g_i)} \rangle | X_m \in X]
% \end{align*}

% As for Vertical Augmentation, a single \textit{Activity} is also created, which connects \textit{Used} and \textit{Generated} entities through \textit{Used} and \textit{wasGeneratedBy} relationships between each pair of entities representing data in the same column, that is, where \textbf{var:F} = \textbf{var:F'}, and an additional \textit{wasDerivedFrom} relationship is also added.

\subsubsection{Data transformation} \label{sec:data-xform}  $\tau_{f(X)}$ takes features $X \subset S$ and computes derived values, which are used to update elements of $D$, but without generating new elements.
The bindings reflect such in-place update, but as the new value for each element is defined by $f()$, we assume for simplicity that \textit{all} values are updated, although, in reality, some will stay the same, as shown for instance in Ex.~\ref{ex:imputation} (imputation). The resulting bindings reflect this many-many relationship, where (potentially) all values in a column $X_m \in X$ are used to update (potentially) all values in that same column (and this applies to each column).
Thus, the provenance document consists of $|X|$ provlets, one for each \added[id=pm]{column}, with bindings defined as follows.
\textit{Used} entities:
\[ [\langle F=X_m, V = D_{i,X_m}, I=i\rangle | i:1\dots n] \]
\textit{Generated} entities:
\[
[\langle F' = X_m, V'= f(D_{*,X_m}), J=i \rangle | i:1\dots n]
\]
\textit{Used} and \textit{wasGeneratedBy} relationships, mediated by an \textit{Activity}, are created between each \textit{Generated} entity and all of the \textit{Used} entities having the same $X_m$, along with the corresponding \textit{wasDerivedFrom} relationships.

\added[id=pm]{It is worth clarifying one potential limitation that occurs when the data derivation operator contains parameters whose values are set by inspecting the input dataframe. In our approach, these values are not ``used'' by the operator, despite the fact that, in reality, the operator is input-dependent. As an example, consider a Scaling operator, which scales each value in column $X$ using a range that is defined by the min and max values found in $X$. According to the template just defined, this operator produces a set of 1-1 derivations, namely from each output value in $X$, back to its corresponding input value. However, in the current approach the fact that the Scaler depends on the input values of $X$, which it has inspected, is not captured.}

\subsubsection{Join} \label{sec:join-template}

In Section~\ref{sec:datamodel} we introduced a join operator: 
$D' = D^L \join^t_{C} D^R$ where condition $C$ may involve any columns $F \subset S^L \cup S^R$.
As an example, consider $S^L = [A, B, C]$, $S^R = [A, C, D, E]$ and $C \equiv D^L.A = D^R.A ~\mathbf{and}~ D^L.B = D^R.D$, thus $F = \{ D^L.A, D^R.A, D^L.B, D^R.D \} $.

\replaced[id=pm]{Let $D^L_i = [x, y, c_1], D^R_j = [x, c_2, y, e] $ be two tuples that contribute a result tuple $D'_h = D^L_i \join^t_{C} D^R_j = [ x, y, c_1, x, c_2, y, e]$.}
{Let $D^L_i, D^R_j$ be two tuples that contribute a result tuple $D'_h$ to $D'$, i.e.,
$ D^L_i \join^t_{C} D^R_j = D'_h$.}

\begin{figure}
    \includegraphics[width=.9\columnwidth]{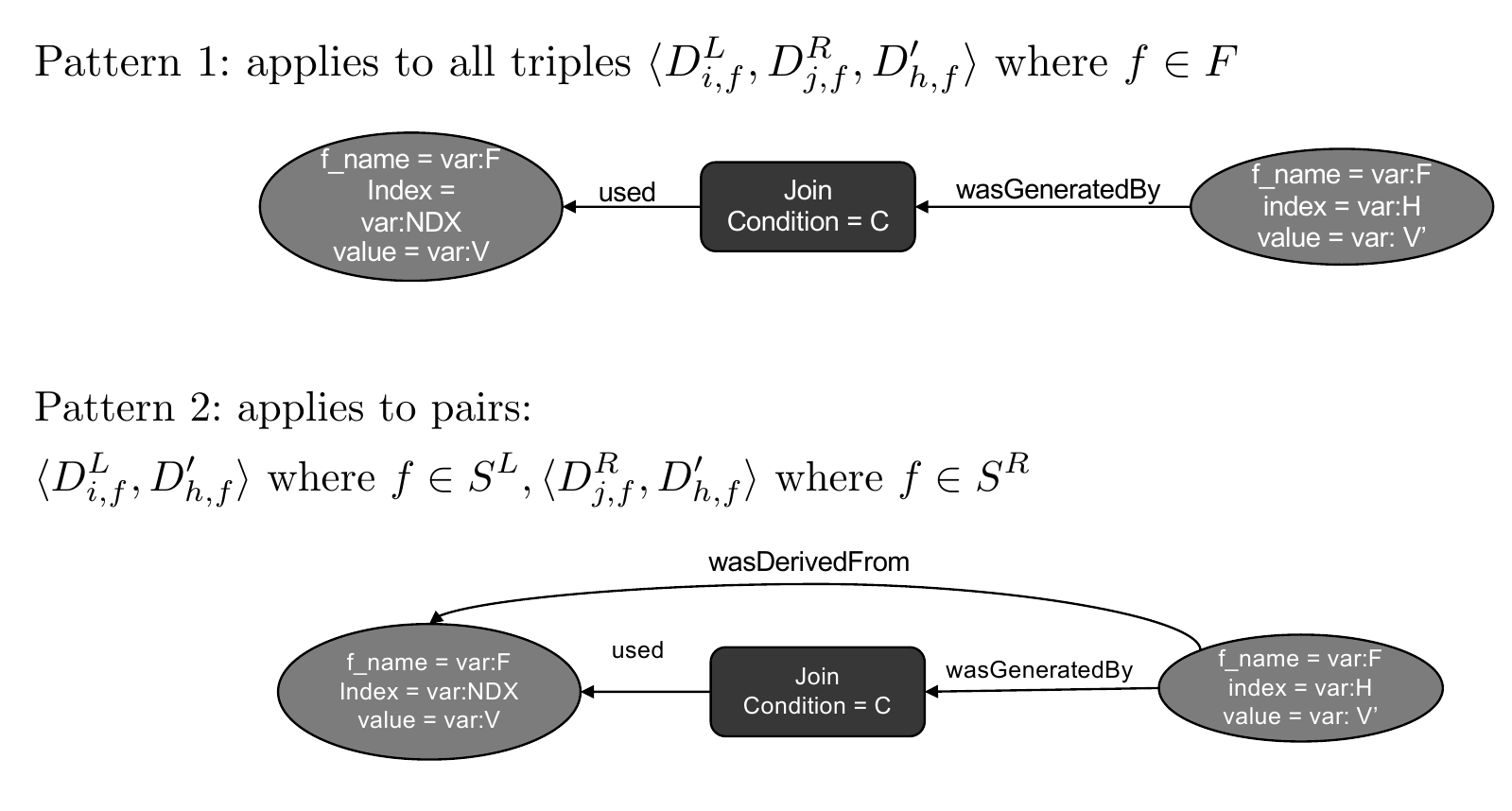}
    \centering
\caption{PROV templates for joins.}
\label{fig:join-patterns}
\end{figure}

Note that \replaced[id=pm]{$D^L_i, D^R_j$}{these two tuples} correspond precisely to the \textit{witness} tuples in the \textit{why-provenance} of $D'_{h,f}$, \added[id=pm]{for some attribute $f \in F$}, as defined in~\cite{buneman_why_2001}. 
The why-provenance of $D'_{h,f}$ can be expressed formally in terms of the two contributing tuples, i.e., using the polynomial notation proposed in~\cite{green2007}. 
However, here we are interested in the more granular derivations at the level of the single values, rather than of the entire tuple.
To express the fine-grained provenance of a value $D'_{h,f}$ in the result, we first consider the values
$D^L_{i,f}$, $D^R_{j,f}$, $f\in F$, 
\textit{used} by the join operator to evaluate $C$:
\[ \mathit{used} =  \{ D^L_{i,f} \cup  D^R_{j,f} | f \in F \} \added[id=pm]{= [D^L_{i,A} = x, D^L_{i,B} = y, D^R_{j,A} = x, D^R_{j,D} = y]}\]
We apply template (1) in Fig.~\ref{fig:join-patterns} to assert that each value in $D'_{h,f}$ \textit{was generated by} the join operator and that the operator \textit{used} all the values in the $\mathit{used}$ set.

This is achieved using the following binding generator:
\begin{align*}
 &\text{for } f\in F: \\
 &\quad\quad \text{if } f \in S^L: \langle F=f, NDX=i, H=h,
  V=D^L_{i,f}, V'=D'_{h,f} \rangle \\
  &\quad \quad \text{if } f \in S^R: \langle F=f, NDX=j, H=h,
  V=D^R_{j,f}, V'=D'_{h,f} \rangle
 \end{align*}

Secondly, we express that each value in the result was \textit{derived from} the corresponding value in one of the two operands, and that the derivation is supported by a \textit{usage/generation} pair as shown in template (2) in Fig.~\ref{fig:join-patterns}. 
Note that this template covers both the case where a feature is used as part of an equijoin condition, such as $A$ in the example, and also the case where null values are generated as part of an outer join.
Template 2 is instantiated using the following bindings generator:

\begin{align*}
 &\text{for } f\in S^L: %\\
 %&\quad \quad 
 \langle F=f, NDX=i, H=h, V=D^L_{i,f}, V'=D'_{h,f} \rangle \\
 &\text{for } f\in S^R: %\\
  %&\quad \quad 
  \langle F=f, NDX=j, H=h, V=D^R_{j,f}, V'=D'_{h,f} \rangle
 \end{align*}

\begin{figure}
    \centering
    \includegraphics[width=\columnwidth]{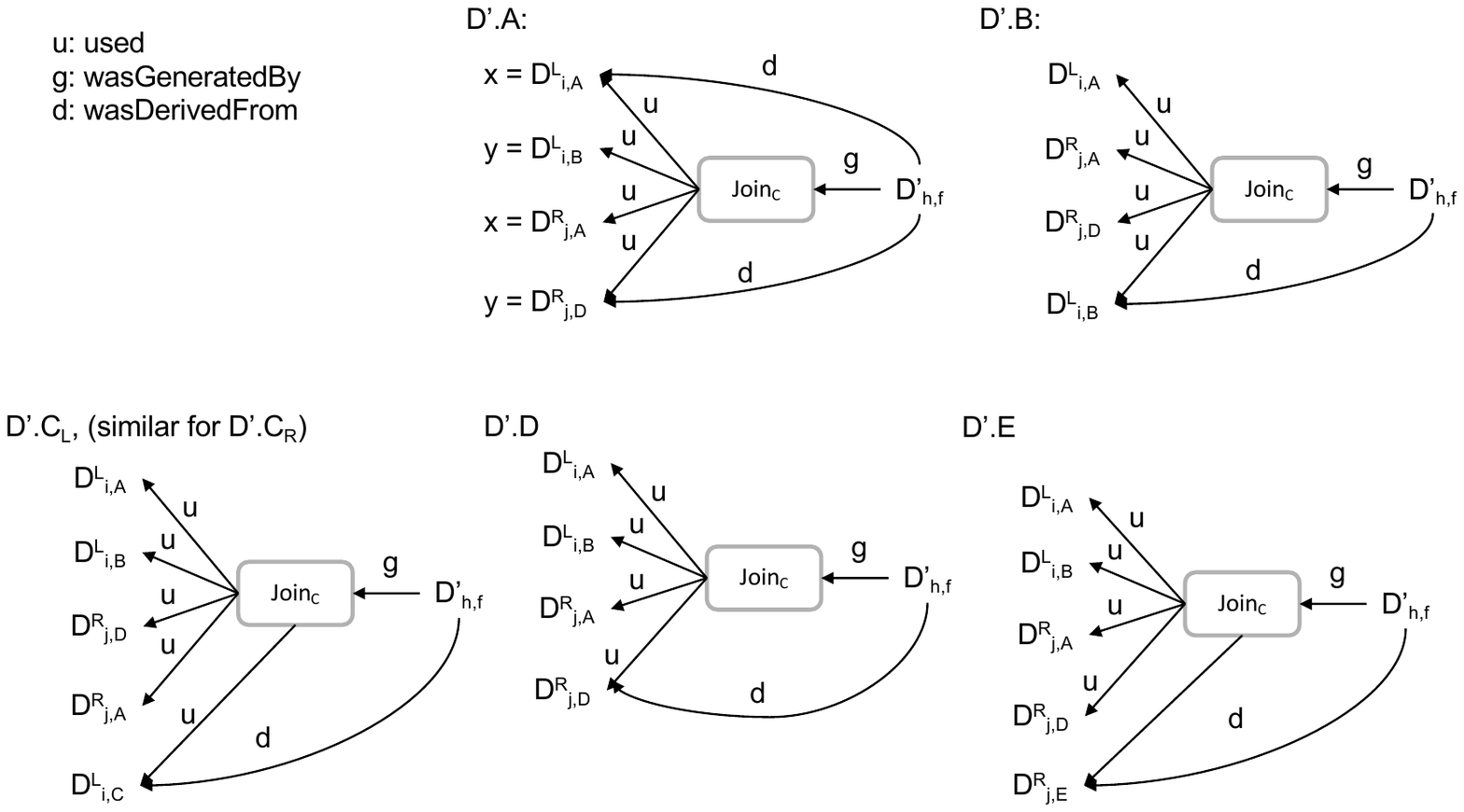}
\caption{Instantiated PROV templates for the example in the text}
\label{fig:join-instance-patterns-example}
\end{figure}

Each of the two templates generates a PROV fragment, and these are then combined by virtue of their common entities and activity (the join operator).
Fig.~\ref{fig:join-instance-patterns-example} \added[id=pm]{(where the actual values of $D'_{h,f}$ are only shown in the first provlet, to avoid overloading the Figure)} shows the provenance fragments for values $D'_{h,f}$ for a generic tuple $h$ and for each $f$.
Note in particular that the generation relationships in templates 1 and 2 do not result in multiple generation arcs in the final provenance, as those have identical source and sink nodes (i.e. the entity representing the value and the activity representing the join).

\subsubsection{Append}
Consider again Example~\ref{ex:fusion} in Section~\ref{sec:datamodel} (page~\pageref{sec:datamodel}), where a dataset $D^L$ with schema $S^L = [\text{CId},  \text{Name}]$ is appended to $D^R$ with schema $S^R = [\text{CId},  \text{Gender},   \text{Age},    \text{Zip},    \text{Name}]$: $D' = D^L \union D^R$.
Let $n,m$ be the number of rows in $D^L, D^R$, respectively. Observing that the order of the rows in the operands is preserved in the result, we identify four types of output values $D'_{i,f}$:
(1) values derived from a corresponding $D^L_{i,f}$, when $i<n_1$ and $f\in S^L$; 
(2) values derived from a corresponding $D^R_{i,f}$, when $i \geq n_1$ and $f\in S^R$; 
(3) Null values when $i<n_1$ and $f\notin S^L$; 
(4) Null values when $i \geq n_1$ and $f\notin S^R$.

\begin{figure}
    \includegraphics[width=.8\columnwidth]{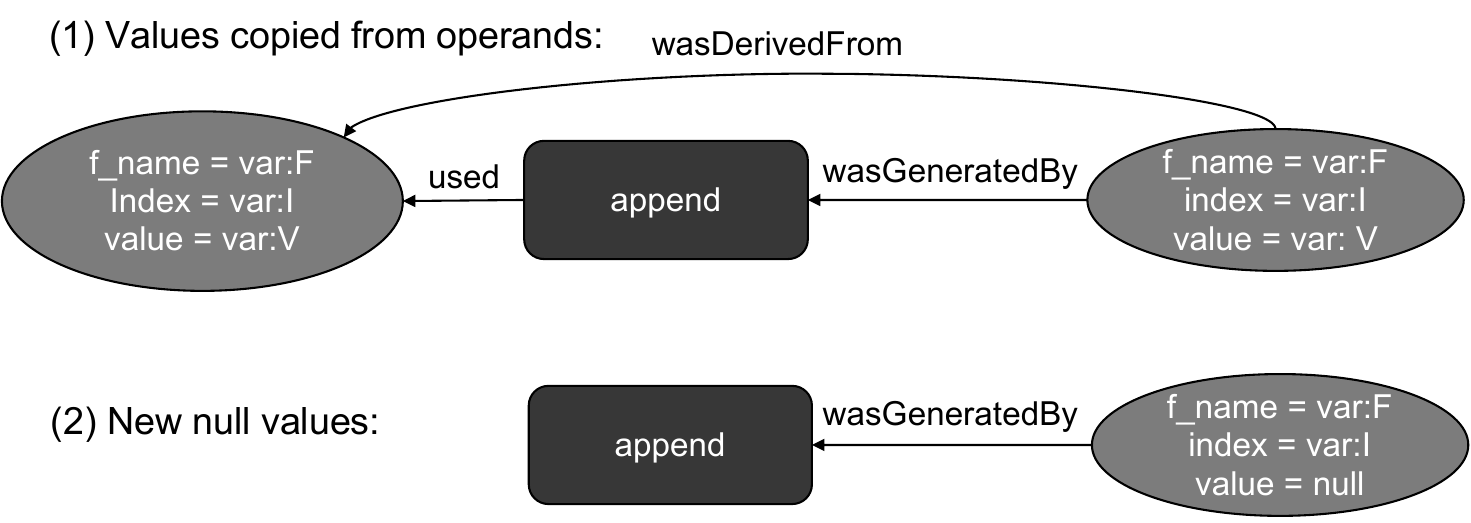}
    \centering
\caption{PROV template for append}
\label{fig:append-template}\end{figure}

A derivation relationship is created for cases (1) and (2), which is supported by a corresponding generation-usage pair of relationships, with the operator as the mediating activity; while for cases (3) and (4), only a generation relationship is created.
Fig.~\ref{fig:append-template} shows the PROV template for this pattern.

The binding generator function for derivations, generation, and usage of copied values is defined as follows:

\begin{align*}
 &\text{for } i:0 \dots n_1 -1: %\\
 %& \quad 
 \text{if } f \in S^L \text{ then }  \langle F=f, NDX=i, V=D'_{i,f} \rangle \text{ \# template (1) applies }\\
&\text{for } i:n_1 \dots n_2 -1: %\\
  %&  \quad 
  \text{if } f \in S^R \text{ then } \langle F=f, NDX=i  \rangle \text{ \# template (2) applies } \\
\end{align*}

\section{Provenance generation} \label{sec:algorithms}

The combination of provenance templates and corresponding binding rules, embodied by the \textit{prov-gen} functions, which we just presented, provide a formal description of the \textit{provenance semantics} associated with each of the core operator classes: data reduction, augmentation,  transformation, and fusion.
In this section, we present a concrete approach to provenance generation that is grounded in this formalisation.

\subsection{The approach}

% \added[id=pm]{CLARIFY THIS IS THE BLACK BOX APPROACH}

Provenance generation operates by (i) observing the execution of operators that consume and generate datasets,  (ii) analysing the value and structural changes between the input(s) and output datasets for that command, and (iii) based on the observed change pattern, select one or more of the templates described in the previous section, to capture the dependencies between the elements of the datasets that have changed.
This approach ensures that the topology of the resulting provenance graph is consistent with the templates, but it also broadens the scope of the operators for which provenance is generated, namely to \textit{any operator} that transforms an input into an output dataset.

As a simple example, consider an imputation operation
%{\small
%\indent \texttt{df1 = df.fillna(0)}}\\
that causes some previously null values to be set to 0, in some (or all) of the columns. This values change pattern is easily recognised and is used to trigger provlet generation using the appropriate template, in this case \textit{data transformation} (cf.~\ref{sec:data-xform}).

More general transformation patterns can be captured using more than one template, and by composing the resulting provlets. For example, consider a pipeline like the following, in which $D_a$, $D_b$, and $D_c$ are the input datasets and $f$ is an imputation function over a feature $K$ of $D_a$:
\[
\begin{array}{cl}
    & D_1=\tau_{f(K)}(D_a)\\
    & D_2=D_b \join^{\tt outer}_{K_1=K_2} D_c\\
    & D_3=D_1 \union D_2
\end{array}
\]
%
%\small 
%\indent \texttt{df1 = df.fillna(0)} \\
%\indent \texttt{df2 = pd.merge(dfA, dfB, on=['key1', 'key2'])}\\
%\indent \texttt{df3 = df1.append(df2)}\\
%\normalsize
%
Its execution results in a collection of three provlets, each accounting for the dependencies between elements of the datasets (1) $D_1$ and $D_a$; (2) $D_2$ and $D_b$, $D_c$; and (3) $D_3$ and $D_1$, $D_2$, respectively.
At the end of the execution, these provlets are consolidated into a single, final provenance document that accounts for all transformations across the entire pipeline. In this example, this will create a graph of dependencies where elements of $D_3$ are linked through derivation relationships to elements of $D_a, D_b, D_c$.

In the cases above, the change analysis identifies the appropriate template without the need for syntactic analysis of the source code.
In particular, these examples illustrate \textit{simple provenance generation}, so called because provlets are independently generated for each input/output datasets pair. 
More complex \textit{composite provenance generation} can also be achieved, which captures the provenance of an operation implemented by a sequence of commands.
We illustrate this in the next Section for the case of one-hot encoding transformation.

\subsection{Change analysis algorithm}
\label{sec:change-algorithm}

\begin{figure*}
    \centering
    \includegraphics[width=.85\columnwidth]{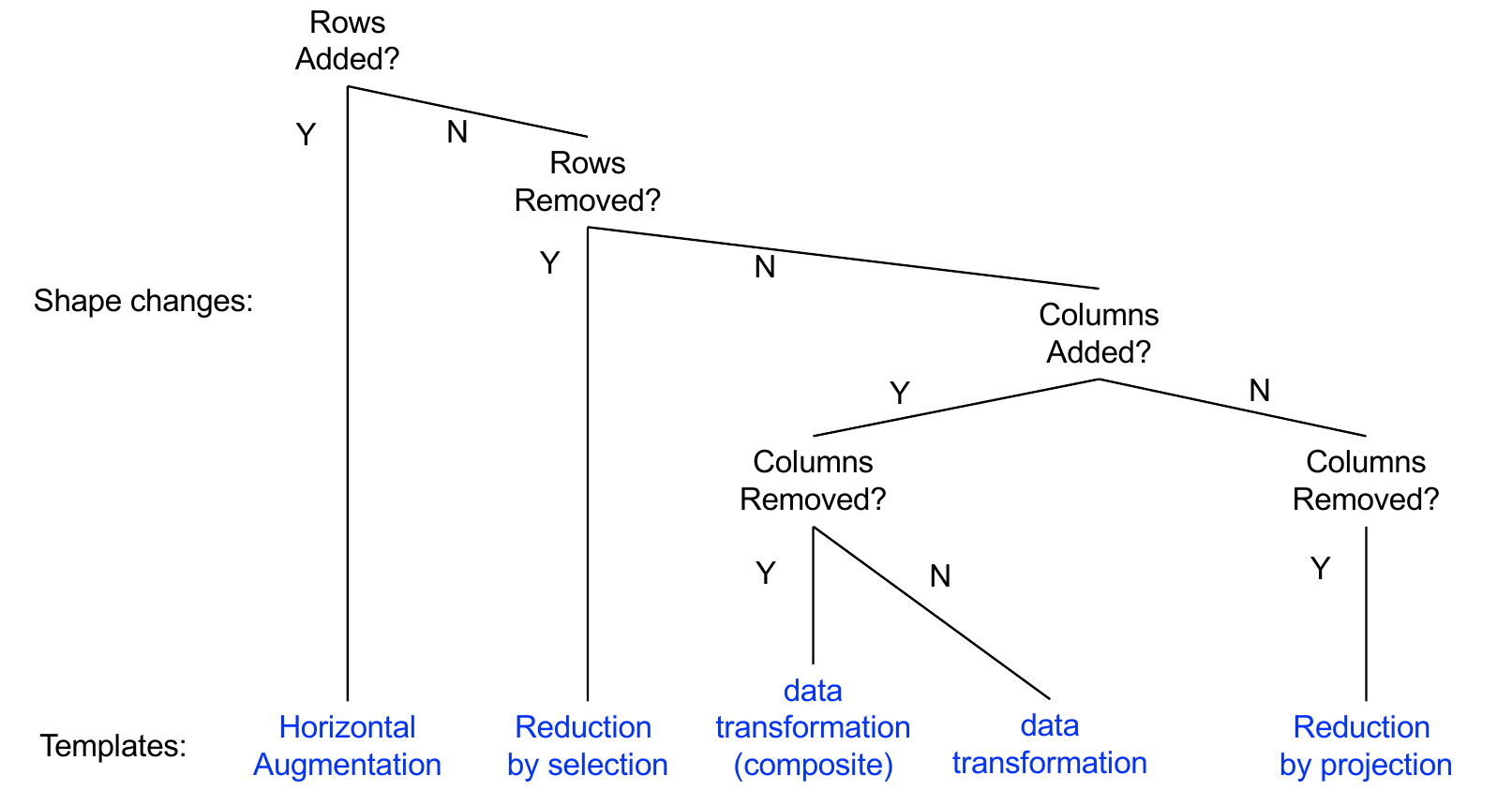}%\\[3mm] 
    \caption{Template selection  for shape change.}
    \label{fig:algorithmPic1}
\end{figure*}

\begin{figure*}
     \includegraphics[width=.7\columnwidth]{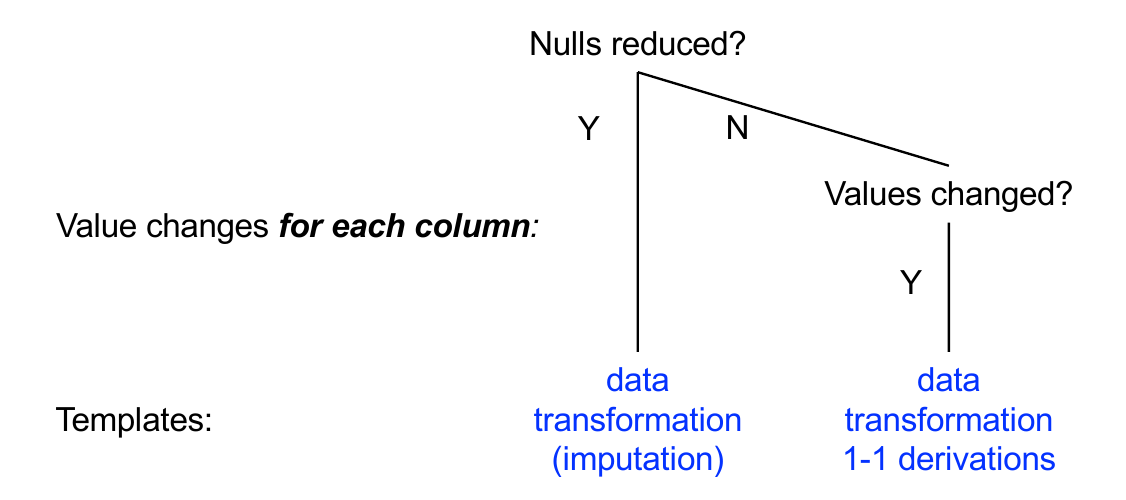}
    \caption{Template selection for value change.}
    \label{fig:algorithmPic2}
\end{figure*}

We now present the dataset change analysis algorithm that is responsible for generating each of the provlets.
The algorithm considers unary and binary operators separately, with help from lightweight code instrumentation. In the following, we only discuss the case of unary operators, as a complete example of join and append provenance has been provided earlier. 
Implementing join provenance efficiently presents new challenges, however, and these are discussed separately below (Section~\ref{sec:joinImpl}).
Details of the code instrumentation required to support provenance generation are provided in the next Section, along with details of the \textit{Observer} pattern \cite{blount2021observed} used to monitor changes in datasets through execution.
The algorithm 
%expects one-input / one-output transformations by default, that is unless code annotations specify operands of a join or append operation. 
%It 
looks at changes in either shape or values between the input and output datasets, denoted $D$ and $D'$, respectively.
The cases listed below are summarised in Figures~\ref{fig:algorithmPic1} and \ref{fig:algorithmPic2}.
Shape changes are detected simply by comparing the number of rows $m, m'$ or the number of columns $n, n'$ in $D, D'$. Value changes are detected by reviewing values within each column.

\paragraph{Shape changes.}

When $m'<m$ or $m'>m$, the horizontal augmentation (cf.~\ref{sec:ha}) or reduction by selection  (cf.~\ref{sec:data-reduction-selection}) templates are applied, respectively.
Adding columns is interpreted similarly, i.e., $n'>n$ triggers the application of the Vertical Augmentation template (\ref{sec:va}). 
However, this condition also causes a list to be created which contains the added columns. 
If $D'$ is then used as input to the next command, producing $D''$, and it is the case that $n''<n'$ when the list of added columns is non-empty, then this is interpreted as a sequence where first a set of columns are added, and then other columns are removed. 
This enables the provenance generator to infer dependencies between such columns. Derivation relationships are thus added accordingly to the provlet that represents the provenance for $D$, 
$D'$, and $D''$.

This composite behavior makes it possible to detect patterns like one-hot encoding, which both adds and removes columns but does so using more than one operator as shown in Figure \ref{fig:one-hot-example}.

%%%%%%%%%%%%%%%%%%%
%%%%% one hot encoding example
The following sequence of operations, in which an input dataset $D$ is first extended by encoding with the function $h$ the values occurring in the feature $B$ and then the such feature is deleted, is routinely used to achieve the result.
\begin{example}[one-hot encoding] \label{ex:one-hot}
\[
\begin{array}{cl}
    & D_1=\horaug_{h(B)}(D)\\
    & D_2=\pi_{\{\it A\ and\ the\ features\ not\ occurring\ in\ D\}}(D_1)\\
\end{array}
\]
%
%\begin{example}[one-hot encoding] \label{ex:one-hot}\\
% \noindent \fbox{
%\begin{minipage}{\columnwidth}
%\vspace{5pt}
%\small
% \texttt{c = 'B'}\\
% \texttt{dummies = []}\\
% \texttt{dummies.append(pd.get\_dummies(df[c]))}\\
% \texttt{df\_dummies = pd.concat(dummies, axis=1)}\\
% \texttt{df = pd.concat((df, df\_dummies), axis=1) \quad \# (1)}\\
% \texttt{df = df.drop([c], axis=1)  \quad \# (2)}
%\vspace{5pt}
%\end{minipage}
% }
%\end{example}
%

\begin{figure}
    \centering
    \includegraphics[width=.6\columnwidth]{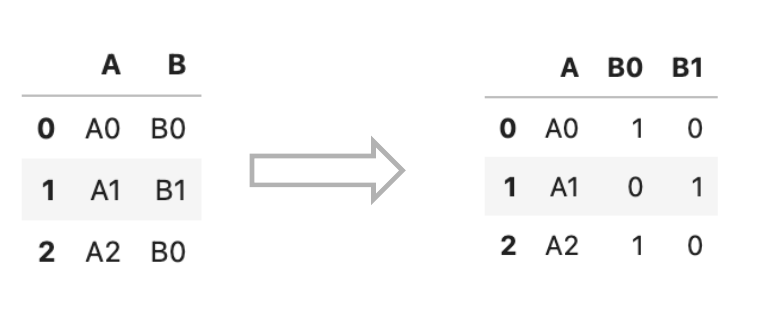} 
    \caption{One-hot encoding transformation}
    \label{fig:one-hot-example}
\end{figure}
\end{example}

After the first operation, the generator would only know that a number of columns have been added (one for each value in column $B$) but would be unable to determine any other dependencies. A list is created with these column names, which will only exist within the scope of the next command.
Executing the second operation results in column $B$ being removed. 
Rather than two connected provlets, here a single provlet is generated, which  accounts for the change in dataset structure, and where derivation relationships are added between each new  column and column $B$ (which is then itself invalidated in the provlet). 
In Fig.~\ref{fig:algorithmPic1} we refer to this as the \textit{composite data transformation} template.

% The list of columns is reset at this point.
% Notice that the flag may not need to be used. For instance a command that simply adds one new column to $D$, followed by other unrelated commands, results in the flag being reset and no derivation edges being created. The standard \textit{data transformation} template is used in this case.

% \rtnote{What flag? Never mentioned before.}

\begin{example}[Provenance of one-hot encoding]\label{ex:one-hot-prov}
Consider the transformation in Fig.~\ref{fig:one-hot-example} implemented by the operations in Ex.~\ref{ex:one-hot}. 
After execution of the first operation, a \textit{VerticalAugmentation} activity is created to account for the \textit{generation} of the new features.
However at this stage, we do not know which elements of the input have been \textit{used}, thus we are also unable to add \textit{derivation} relationships.
After executing the second operation, feature $B$ has been removed, and in the provlet this is recorded by introducing a new  \textit{ConditionalProjection} activity that \textit{invalidates} $B$. Additionally, however, the composite variant of data transformation mentioned above is applied, resulting in a new relationship:\\
\indent \textit{VerticalAugmentation} \textit{used} \texttt{B}\\
as well as derivations:\\
\indent  \texttt{B0} \textit{wasDerivedFrom} \texttt{B}, \\
\indent \texttt{B1} \textit{wasDerivedFrom} \texttt{B}.

\noindent The complete provlet includes the statements:\\
% \begin{minipage}{\columnwidth}
% \vspace{10pt}
\indent \texttt{B} \textit{wasInvalidatedBy} \textit{ConditionalProjection} \\
\indent \texttt{B0} \textit{wasDerivedFrom} \texttt{B} \\
\indent \texttt{B1} \textit{wasDerivedFrom} \texttt{B}
% \end{minipage}
\end{example}

\paragraph{Value changes.} 

This analysis considers one column at a time. 
If some or all of the values have changed in a column $C$, the \textit{data transformation} template is  applied,  with the assumption that there is a one-to-one dependency between each new value $d'_{ij}$ od $D'$ and the corresponding original value $d_{ij}$ of $D$, and this is mediated by the function represented by the operator, as described in Section~\ref{sec:data-xform}.
The case of value imputation is handled separately. 
This is detected simply by comparing the number of null values in $D$ (identified by \texttt{NaN}) to those in $D'$. 
Imputation has occurred when the nulls have been reduced. 
In this case, the \textit{data transformation} template is used (\ref{sec:data-xform}), where by default each value in an imputed column $C$ in $D'$ are assumed to be derived from all values in the same column $C$ in $D$. 
Notice that the generator does not have further information to make the derivation more granular.
Also, when multiple columns have been imputed, each of those is considered independently from the others. This may miss derivations, again for lack of information. For instance, using the MICE algorithm~\cite{MICE} will impute multiple columns, where each new value is derived from values in multiple source columns. This generalisation is not captured by the algorithm.

\subsection{Benefits and limitations of the change analysis approach}
\label{sec:ben-lim}

The approach of using dataset change as the trigger to choose the provenance template and to apply and generate provenance information \replaced[id=pm]{has two main advantages}{provides two inherent capabilities}. 
\added[id=pm]{Firstly, it makes it possible to capture provenance when the internal logic of the operators is not accessible to the observer. This has been referred to as the ``black-box'' problem by the provenance community~\cite{DBLP:journals/corr/abs-1201-0231}.}
\added[id=pm]{Secondly, it enables capturing the provenance of operator compositions. In the presentation of
this work, we mainly describe the provenance generated for a single operator execution.
However, by looking only at dataset change, we can allow multiple operators to execute
and generate the provenance record for this group of operators. An example is the “stateful”
shape change analysis above, which keeps track of data transformations across more than
%
% \removed[id=pm]{
one operator, in order to accurately infer derivation dependencies.}

\added[id=pm]{One limitation that is intrinsic to this approach is in complex cases such as when UDFs are employed. In this case, while the algorithm can detect which tuples have changed, it cannot identify which inputs caused the change, thus it must assume that all inputs were used by default.}
% \begin{enumerate}
%     \item Allows capture of provenance for operators not specified in advance. By mapping the effects of an operator, as opposed to the logic of the operator, we can map these operators to a transformation type. \replaced[id=pm]{This This effectively eliminates the ``black-box'' vs ``white-box'' problem, where the former refers to operators whose internal logic is not accessible to the observer
    
%     that has plagued the provenance community for decades. 
%     \item  Allows capture of provenance at any level of operator composition. In the presentation of this work, we mainly describe the provenance generated for a single operator execution. However, by looking only at dataset change, we can allow multiple operators to execute and generate the provenance record for this group of operators. 
%     An example is the ``stateful'' shape change analysis above, which keeps track of data transformations across more than one operator, in order to accurately infer derivation dependencies.
% \end{enumerate}}
%
The ability to group operators is beneficial for many reasons. Provenance is often unwieldy, capturing interactions and relationships meaningless for later use. Past works utilize variations in ``Composite'' to help with various tasks. ZOOM \cite{cohen2008}  used the concept of composite step-classes to develop a notion of user views, allowing a user to more easily view and understand a provenance graph. More recently, Ursprung \cite{rupprecht2020} contains provenance at different composite levels based on the capture mechanism able to be deployed in a given situation. In our work, the developer can choose a composite that is correct for their ultimate end needs by having the provenance observer wait for other commands to complete and only look at the final dataset.

\section{Implementation and Architecture} \label{sec:architecture}

In this section, we provide details on (i)~the data architecture used in the implementation, (ii)~the code instrumentation required for the provenance generator to operate, and (iii)~the efficient implementation of provenance capture and provlet composition.

\subsection{System architecture}
\label{sec:arch}

\begin{figure}[ht]
    \includegraphics[width=.6\columnwidth]{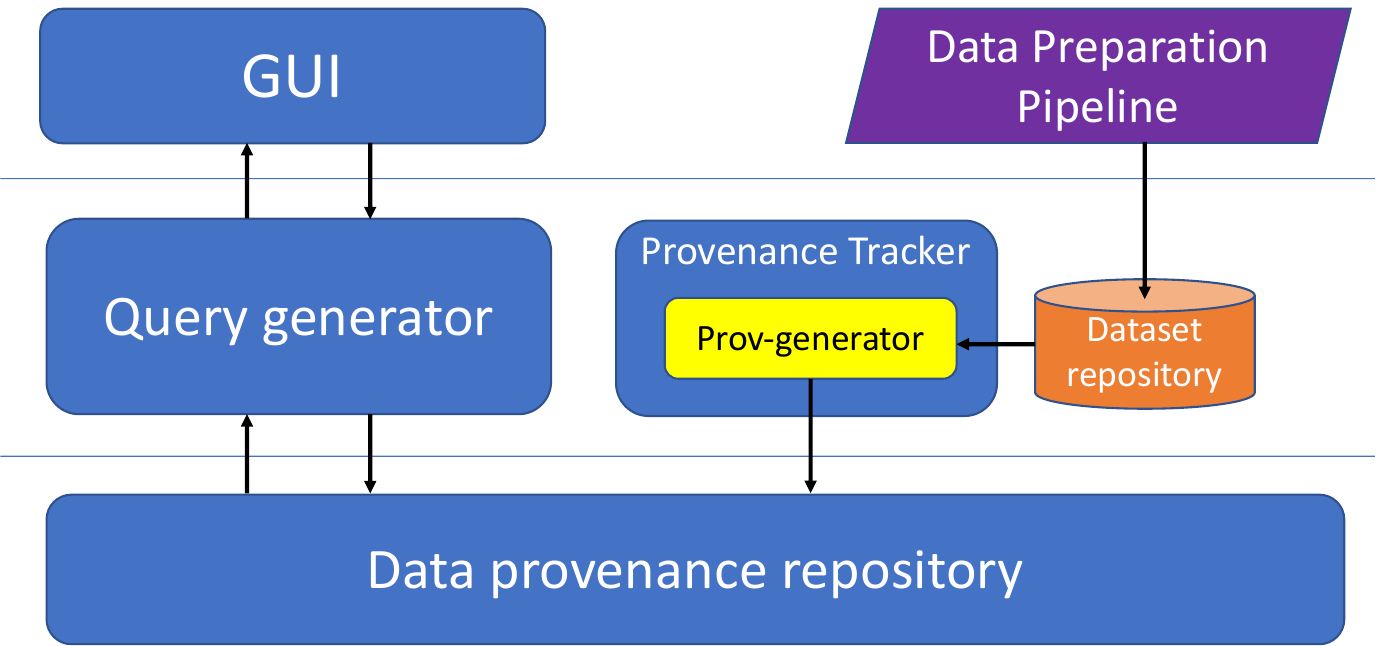}
    \centering
\caption{System architecture}
\label{fig:arch}
\end{figure}

We have \added[id=ac]{created a reference implementation of} the approach to provenance generation illustrated in Section~\ref{sec:algorithms} using the pandas/python library, representing datasets as pandas dataframes\footnote{\url{https://pandas.pydata.org/}}.
The overall architecture for provenance capture, storage, query, and visualisation is shown in Fig.~\ref{fig:arch}. 
The Provenance-Tracker automates the process of detecting and tracking the
provenance of a user-defined pipeline of data preparation. It includes a Prov-generator that produces the provenance of each operator in the pipeline by analyzing its effect on the underlying dataset. This is done at execution time by: (i) identifying the operator under execution on the basis of a series of comparisons between the input and the output datasets, 
(ii) executing the prov-gen function of the core operation that captures the identified operator by suitably instantiating the function template, and 
(iii) storing the provenance data produced by the prov-gen function on an underlying repository. 
Since provenance data have a natural graphical representation, Neo4j, a world-leading, industry-grade, scalable graph database management system, is used for this purpose. \added[id=pm]{Note that while provenance graphs are written to the database at runtime, i.e., while the script is executing, those writes can happen asynchronously, as the graph will only be queried ``post mortem'' after the script has finished executing. This also removes the need to consider a high-performance back end such as an in-memory database.}

The \textsf{Query-generator} allows the user to perform several types of analyses of the data provenance collected for a given data preparation pipeline, by translating a specific data-provenance exploration chosen from a menu of a graphical interface into a query expressed in Cypher, the query language of Neo4j, \added[id=rt]{as it will be illustrated in Section \ref{sec:provexp}}.

\added[id=ac]{
With the current reference implementation we have made several upgrades to the earlier version \cite{chapman2020anonymized}: (1) the data representation format; (2) the storage method, as we now use the Neo4J graph database to store the final provenance graph natively in contrast to \cite{chapman2020anonymized}, where all provenance was serialized in PROV-JSON \cite{w3c-prov-dm,provw3c} (an interoperability format for the PROV data model); and (3)~\textit{observing} provenance from dataframes instead of specifically coding each pandas operator.
}
 \added[id=pm]{We have chosen to represent provenance graph using the standard PROV data model, to ensure some degree of interoperability across applications that want to use the provenance graphs. However, we are also aware that the standard PROV serialisations documented as part of the W3C specification    are not concerned with space utilisation and query performance.
 Thus, our prototype implementation aims to strike a balance between performance and interoperability goals. 
In particular, all framework elements that have changed through an operator are materialised in the provenance (but no new entities are introduced to represent data items that have not changed). 
As all required entities are manifested in the graph, new provenance queries can be written simply using standard Cypher, with minimal knowledge of the internal representation. In a future implementation, one may introduce entities that represent entire tuples or columns, but with the understanding that queries must be aware of these optimisations.}

\subsection{Code instrumentation} \label{sec:code-instrumentation}

\added[id=ac]{
A number of different approaches for capturing provenance from a running process have been documented in the literature. These range from  intentionally placing capture calls within the notebook, to utilizing libraries to compare dataframes for automatic detection, to engaging interactively with the user. A key distinction concerns how much burden can be placed upon the user.
Works such as \cite{allen2010capturing} or \cite{pimentel2017noworkflow} insist on no-human involvement, while others believe that users should be invested in the process of improving their scripts, specifically \cite{lerner2023making} or \cite{zhang2017using} allow  users to enter provenance capture calls at appropriate places. Other contemporary systems, such as MLInspect \cite{grafberger2021}, require the development of specialised add-ons to the code (using a visitor pattern) to create observers. The level of developer involvement is still an open research question for the data science community. }

\added[id=rt]{In this work we aim to implement the strategy described in Section~\ref{sec:change-algorithm} while minimising user intervention.
This is achieved  using an \texttt{Observer} software pattern that acts as a wrapper for dataframes   and relies on a \texttt{Provenance Tracker} object for deriving the provenance of a transformation based on the inspection of the dataframes in input and output.
From an operational standpoint, the Provenance Tracker is equipped with a \texttt{subscribe()} function that allows users to subscribe to one or multiple dataframes for tracking their provenance. 
Basically, the invocation of this function returns the corresponding wrapped dataframes as objects that encapsulate nearly all of the methods inherited from the pandas DataFrame class, enabling provenance generation  in a transparent way during dataframe transformations.
%known as 'TrackedDataFrame,'. These 'TrackedDataFrame' objects encapsulate nearly all of the methods inherited from the pandas DataFrame class, enabling the capture of provenance information during various dataframe transformations.
%
% The basic instrumentation pattern is limited to the following:}
The only required instrumentation is the following:}
\medskip

\begin{center}
\noindent
\fbox{
\begin{minipage}{.7\columnwidth}
\vspace{5pt}
\small
\added[id=rt]{
\texttt{tracker = ProvenanceTracker()}\\
\texttt{df, df2 = tracker.subscribe([df, df2])}
}
\vspace{5pt}
\end{minipage}}
\end{center}

\medskip

\added[id=rt]{
After this, the signature and syntax of methods that operate on a dataframe remain unchanged, as in the examples that follow. However, they now operate on the wrapped dataframes and invoke the internal provenance-capture functionality through the Provenance Tracker.
}

\medskip

\begin{center}
\noindent
\fbox{
\begin{minipage}{.7\columnwidth}
\vspace{5pt}
\small
\added[id=rt]{
\texttt{\# Imputation}\\
\texttt{df = df.fillna('Imputation')}\\
\texttt{\# Feature transformation of column D}\\
\texttt{df['D'] = df['D'].apply(lambda x: x * 2)}
}
\vspace{5pt}
\end{minipage}
}
\end{center}

\medskip

%\added[id=rt]{
%In the previous example, it was shown how it is possible in order to capture the provenance related to both imputation and feature transformation activities. It is now possible to capture both activities with a single instruction, as in the following example:
%}

%\medskip

%\begin{center}
%\noindent
%\fbox{
%\begin{minipage}{.7\columnwidth}
%\vspace{5pt}
%\small
%\added[id=rt]{
%\texttt{\# Imputation and feature transformation of column D}\\
%\texttt{df = df.applymap(lambda x: 5 if pd.isnull(x) else x * 2)}
%}
%\vspace{5pt}
%\end{minipage}
%}
%\end{center}

%\medskip

\added[id=rt]{
Similarly, provenance generation for the join operation can be done without the need to invoke additional auxiliary functions, as follows.
}

\medskip

\begin{center}
\noindent
\fbox{
\begin{minipage}{.7\columnwidth}
\vspace{5pt}
\small
\added[id=rt]{
\texttt{df = df.merge(right=df2, on=['key1','key2'], how='left')}
}
\vspace{5pt}
\end{minipage}
}
\end{center}

\medskip

\added[id=rt]{
%Depending on the input dataframe, this could generate imputation for some columns and feature transformation for others.
The activity of the Provenance Tracker can be temporarily disabled 
to capture the provenance of an operation made of several basic data transformations. This is done by using the \texttt{dataframe\_tracking}  property as in the example that follows, which implements the provenance capture of the one-hot encoding sequence illustrated in Example \ref{ex:one-hot}.}

\medskip

\begin{center}
\noindent
\fbox{
\begin{minipage}{.7\columnwidth}
\vspace{5pt}
\small
\added[id=rt]{
\texttt{tracker.dataframe\_tracking = false}\\
%\texttt{c = 'B'}\\
\texttt{dummies = pd.get\_dummies(df['B'])}\\
\texttt{df = df.concat(dummies.add\_prefix('B'+'\_'))}\\
\texttt{tracker.dataframe\_tracking = true}\\
\texttt{df = df.drop([c], axis=1)}
}
\vspace{5pt}
\end{minipage}
}
\end{center}

\medskip

\added[id=rt]{
In this example, the changes made by the horizontal augmentation on the original dataframe are produced but taken into account only during the subsequent operation, when the \texttt{dataframe\_tracking} property is set to `true'.
}

\deleted[id=rt]{
The case of joins and append commands requires additional instrumentation because the tracker needs to observe two operands. It also needs to be told which keys are used in the join operation, in order to correctly track dependencies from key values to non-key values, as described later in this Section.
The following fragment implements a tracked (left) join between \texttt{df1} and \texttt{df2} using keys \texttt{key1} and \texttt{key2}:
}

\deleted[id=rt]{
The pattern for Append is similar, and examples are omitted.
}

\subsection{Composing provlets into a complete provenance document}
\label{sec:generation}

A complete provenance document is produced by combining  the collection of provlets that results from each instance of change analysis. Specifically, one provlet is generated for every transformation and every element in the dataframe that is affected by that transformation.
The final document is composed of such a collection of provlets, where entity identifiers match across provlets, and never needs to be fully materialised, as explained shortly. 

\begin{figure}
    \includegraphics[width=.8\columnwidth]{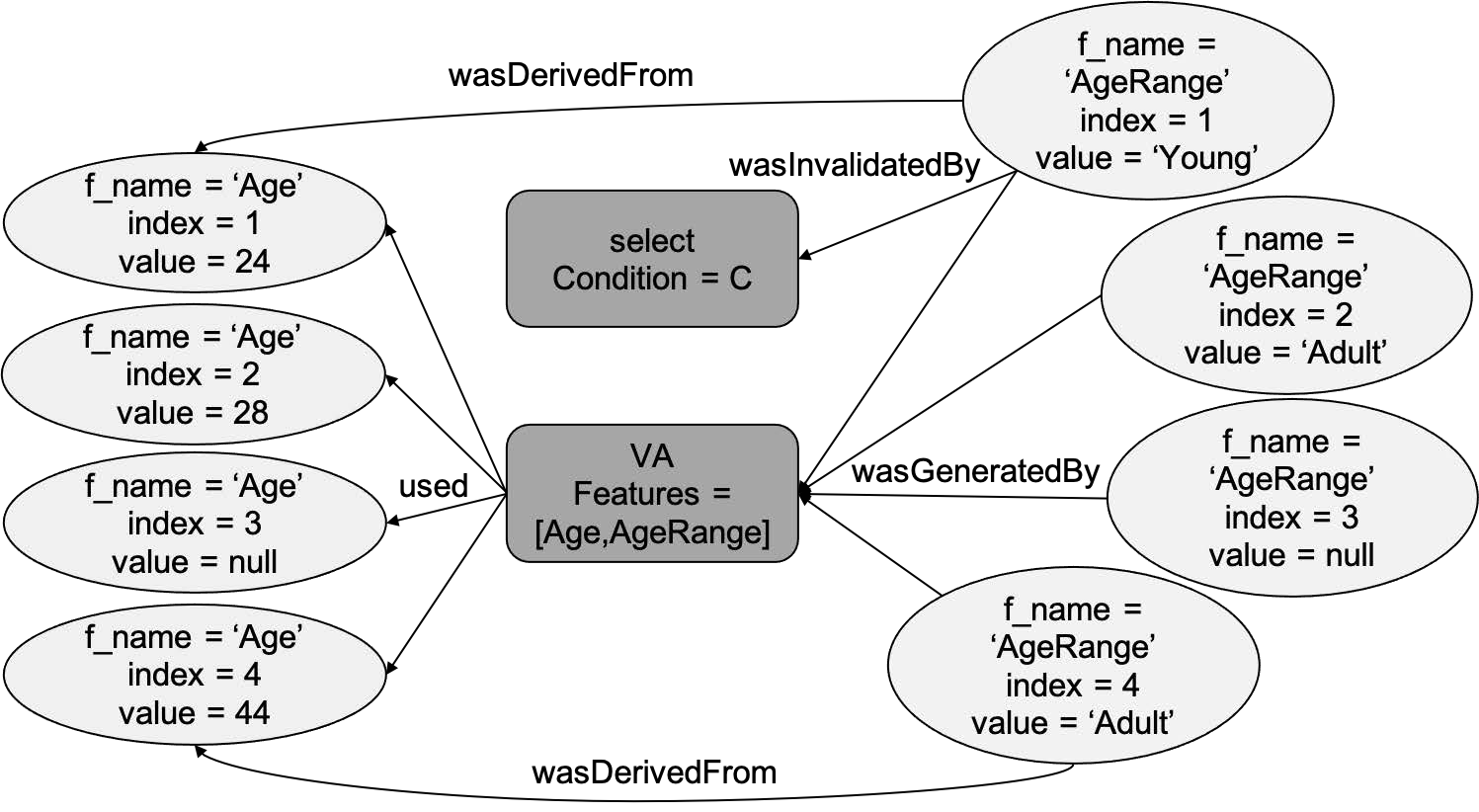}
    \centering
%\caption{Example of provlet composition.}
\caption{Provlet composition.}
\label{fig:example-composition}\end{figure}

Consider for instance the following pipeline:
\[ 
\sigma_C(\horaug_{f_1(\ft{Age}): \ft{ageRange}}(D))
\]
where $C = \text{\{\ft{AgeRange} $\neq$ \texttt{`Young'}\}}$ and $D$ is the dataset of Example~\ref{ex:augment}.
The corresponding provenance document is represented in Figure~\ref{fig:example-composition}.
Applying vertical augmentation produces one provlet for each record in the input dataframe, showing the derivation from \textbf{Age} to \textbf{AgeRange}. 
The second step, selecting  records for `not young' people, produces the new set of provlets on the right, to indicate invalidation of the first record, as per the template at the bottom of Figure~\ref{fig:all-templates}.
Note that the ``used'' side on the left  refers to existing entities, which are created either in the  pipeline from the input dataset, or by an upstream data generation operator.  

Provlet composition requires looking up the set of entities already produced, whenever a new provlet is added to the document.
One simple way to accomplish this is by eagerly keeping the entire document in memory, along with an index for all entities, and by mapping each entity to the corresponding data element it represents.
While this can be accomplished using readily available Python PROV libraries \cite{provpy}, it does not scale well to the volume of entities required to represent large dataframes in cases where more than a handful of transformation operators are involved.
Instead, we have followed a continual append approach for provenance composition in which each p-gen function generates a set of provlets (in the worst case one for each element in the dataframe) that are just collected in a partial document and stored in the underlying repository. 
This allows the provenance to be collected quickly at the execution of each script, and be assembled later, minimizing execution dependencies and possible bottlenecks during the actual execution of the pipeline.

\subsection{Efficient provenance generation} \label{sec:joinImpl}

The overhead for provenance collection and composition described above can be minimised by using Python's multiprocessing library to parallelise the most expensive operations, observing that (i) dataframes can be split into chunks and provenance entities generated independently for each chunk, and (ii) the provlets generated by each parallel process can be independently written to disk, and then asynchronously inserted into Neo4J. 
Assuming that provenance graphs are only queried after the end of script execution, this provides a scalable back-end solution despite the potential limitations of Neo4J's centralised architecture.

Using parallel processes to write provlets to disk is straightforward, as there are no dependencies amongst these processes.  As an example, for one-hot encoding provenance consisting of about 2M entities, we observe a stable $70\%$ improvement in writing times using 12 processes, relative to a sequential baseline.
In practice, at most one chunk is created for each available CPU thread and allocated to one process.
One slight complication is that assigning each generated entity to its corresponding dataframe element requires keeping track of the relative order of the chunks in the dataframe. This is accomplished using a queue (further details omitted). Unlike for write operations, here performance gains depend on the complexity of the specific operator, i.e., of the template used. Empirical results indicate an average of $60\%$ improvement relative to the sequential baseline.
Performance figures from our comprehensive evaluation are reported in the next section.

Joins present an interesting implementation twist to provenance generation mechanism.
A naive implementation of join provenance that creates instances similar to  template in Section~\ref{sec:join-template}, would simply link each row of the output dataframe to the two input DataFrames using rules to infer the derivations for every single item in a row. Unfortunately, joins expose one of the problems of our approach which looks at the input/output datasets and not the operator itself. Because we are not linked directly to the join operator, which may or may not have the standard guarantees of a database system, re-creating which rows in the input dataframes and their relationship to the output row in the dataframe takes effort.

Consider the naive implementation of creating join provenance records using our data-observation approach. For every row in the output dataframe, the join key(s) must be identified within the data, and the actual data values in the remaining features noted. Then, the input dataframes must be scanned to locate the key(s), and the row examined to determine if it contains the appropriate data values to match the output dataframe row. Initial experiments indicate that the scan operation takes ~0.07s per row.  

% Capturing the provenance of all possible join cases (left, inner, outer, and their variants) is a complex task since the results can vary significantly depending on the type of join used, and on the number of attributes and rows involved. Therefore, we have adopted a special method for the provenance of joins that try to keep low the access time in all possible situations.

%The challenge of tracing from which entity an element of the join derives is that the input data frames is known in advance. In join operations, but especially in the case of the inner join, the resulting dataframe can also be significantly smaller than the input ones and with different indexes.}

\begin{figure} 
      \includegraphics[width=.7\columnwidth]{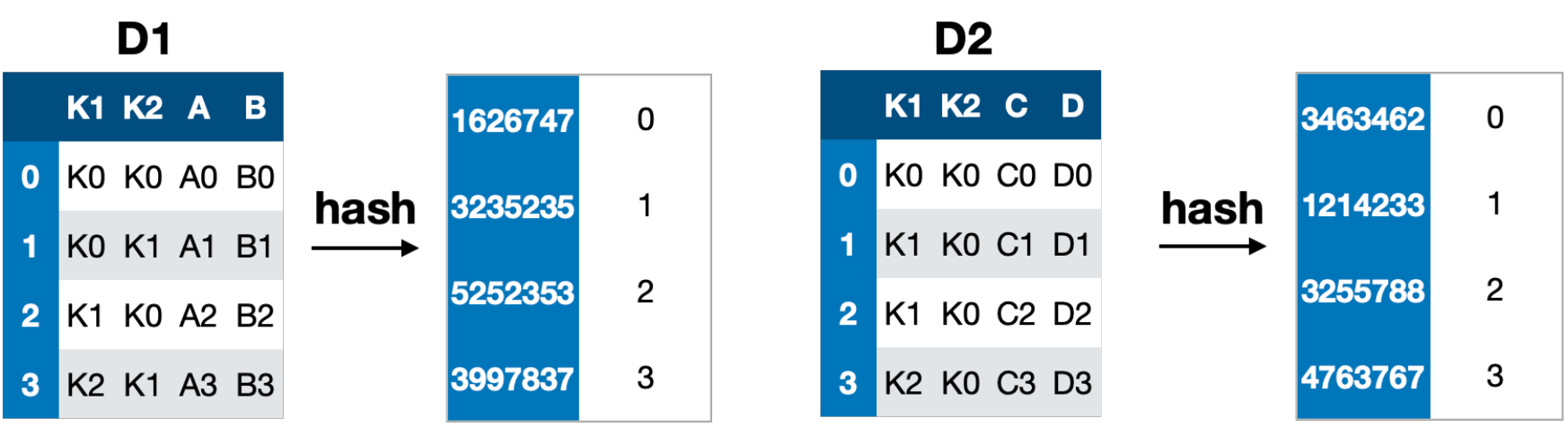}
      \caption{Hashing of the input dataframes}\label{fig:joinin}
      \vspace{2mm}
      \includegraphics[width=.8\columnwidth]{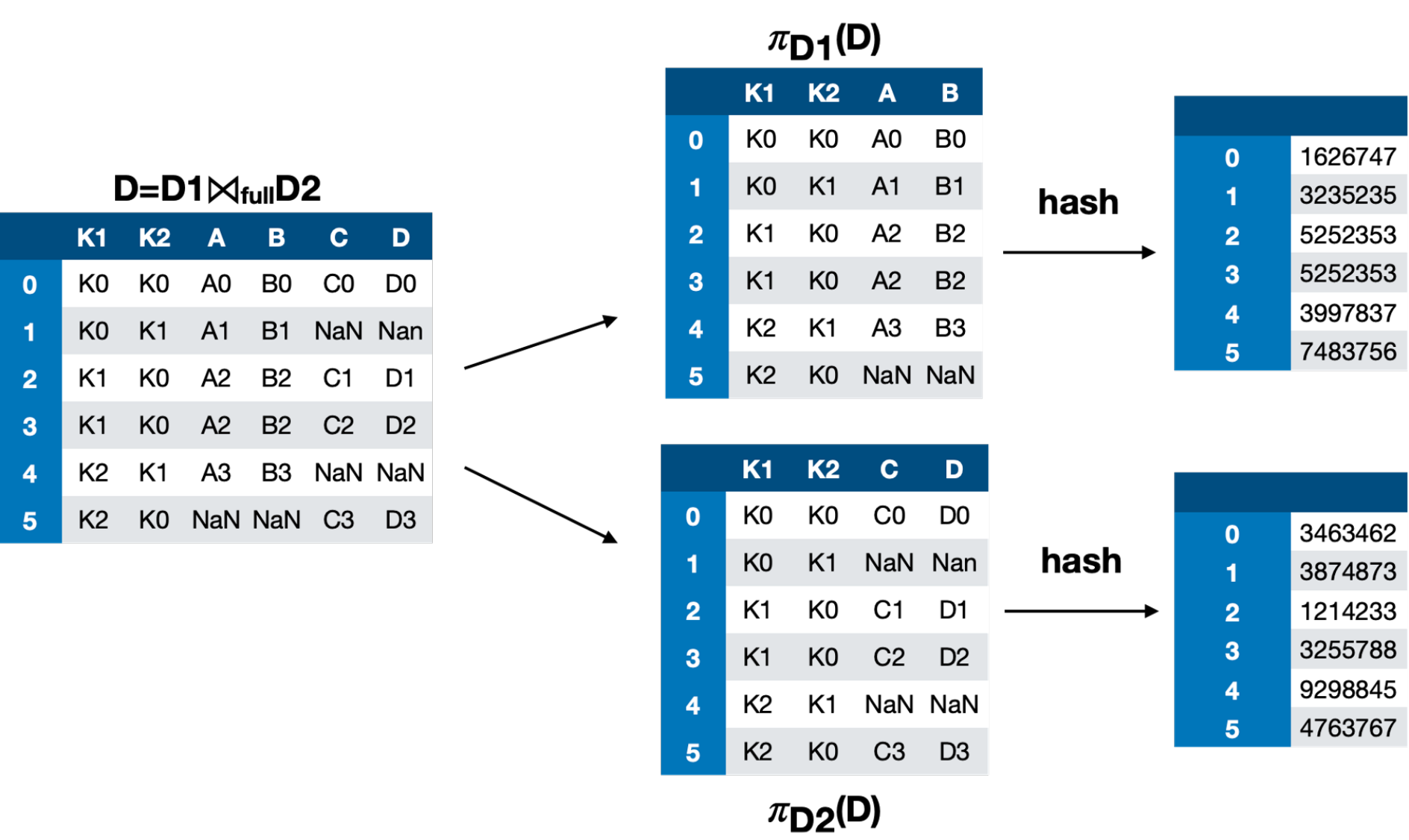}
        \caption{Hashing of output dataframe}\label{fig:preparative-hash}\label{fig:joinout}
        \vspace{5mm}
    \includegraphics[width=.9\columnwidth]{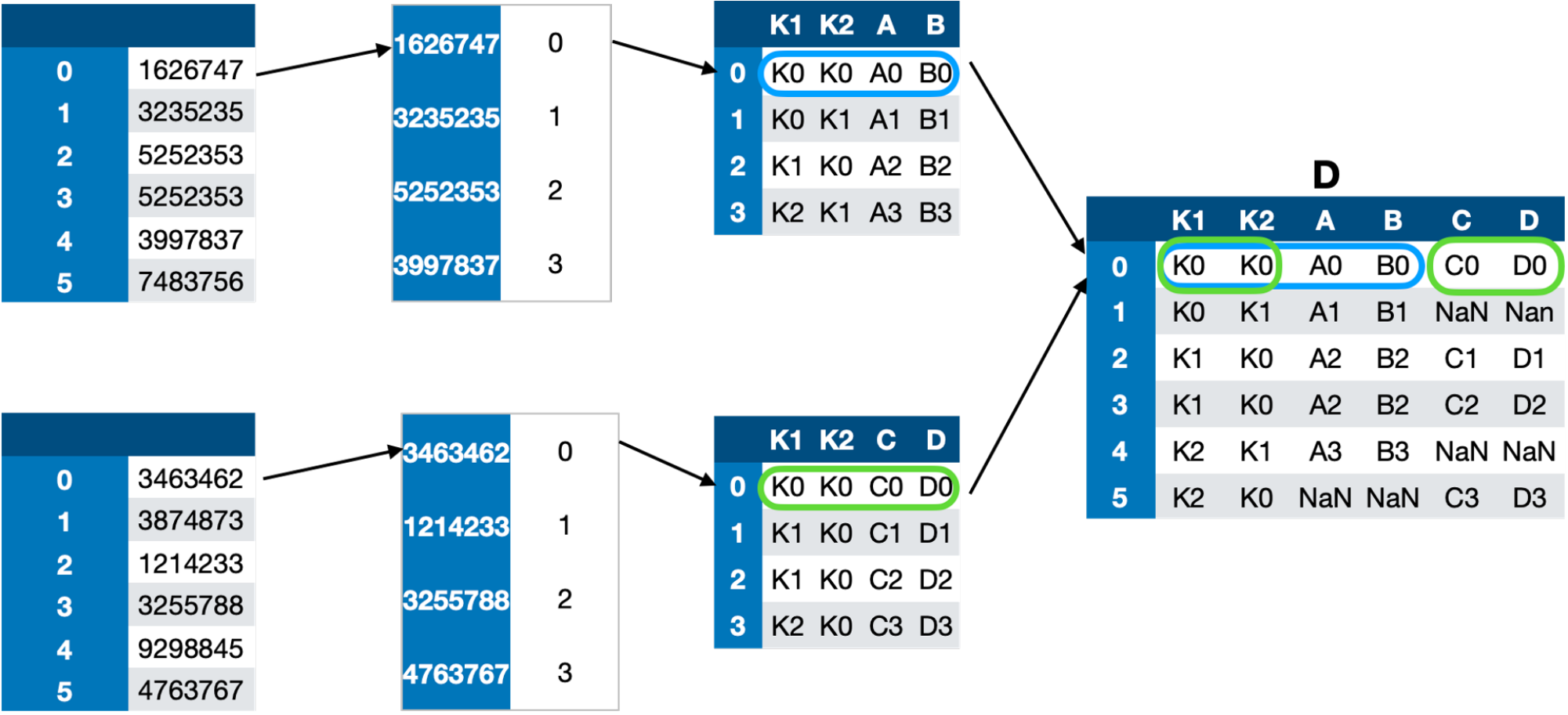}
\caption{Provenance derivation for the join}\label{fig:join}
\label{fig:join-tracking-example}
\end{figure}

To overcome this problem, a more efficient implementation makes use of hash tables. Specifically, two hash tables with the same structure are generated, one for each input dataframe, having, as key, a hash obtained from each row and, as value, the original index of the row (Fig. \ref{fig:joinin}). To derive the provenance, the output dataframe $D$ is then decomposed into two dataframes obtained by projecting $D$ on the columns of the input ones. Then, the two dataframes so obtained are hashed using the same function above (Fig. \ref{fig:joinout}). This allows us to derive easily the provenance of each row of the join as shown in Fig. \ref{fig:join}.

\section{Evaluation} \label{sec:evaluation}
All experiments illustrated in this section were performed on a MacBook Pro with 2.6 GHz Intel Core i7 6 core and 32GB RAM 2400MHz. We focus our evaluation on pandas operators for data cleaning and pre-processing, and can theoretically accommodate ML libraries such as scikit-learn as shown in Table \ref{tab:mloperators} although our reference implementation does not explore using their libraries. 
Given the reference prototype nature of the implementation, the evaluation does not address scalability and performance requirements of a production-grade system.

%\subsection{Utility and Performance in Real World Pipelines} \label{sec:evalrealperf}
\subsection{Analysis with real world pipelines}
\label{sec:evalrealperf}

\myparagraph{Datasets.} 
In Table~\ref{tab:provqueries} reported at page \pageref{tab:provqueries} we have shown classic provenance queries in terms of data input and output. In order to evaluate if we can answer those queries, we have captured data provenance in three real world pipelines involving different types of preprocessing steps. The datasets are described in Table~\ref{tab:datasets}.

\begin{table}[ht]
    \caption{Datasets used for evaluation.}%\vspace*{-3mm}
    \label{tab:datasets}
    \centering\small
    \begin{tabular}{|l|c|c|c|}
        \hline
         & \textbf{German} & \textbf{Compas} & \textbf{Census} \\
         & \textbf{Credit} & \textbf{Score} &  \\
        \hline \hline
        \textbf{Records} & 1000 & 7214 & 32561\\
        \hline
        \textbf{Features} & 21 & 53 & 15 \\
        \hline
        \textbf{\# Operations} & 4 & 7 & 5 \\
        \hline
        \textbf{Output Records} & {1000} & {6907} &  {32561}\\
        \hline
        \textbf{Output Features} & {60} & {8} & {104} \\
        \hline
        \textbf{Provenance Entities} & {85000} & {349970} & {3874264} \\
        \hline
        \textbf{Provenance Activities} & {26} & {7} & {20} \\
        \hline
        \textbf{Provenance Relations} & {255000} & {451412} & {9703396} \\
        \hline
    \end{tabular}
\end{table}

The goal of the \textit{German Credit} pipeline is to predict whether an individual is a good lending candidate. %. Preprocessing steps are as follows: value transformation of 13 distinct columns from codes to interpretable words; generation of two new columns from the column \textit{personal\_status}; the column \textit{personal\_status} was deleted; 11 categorical column were OneHot encoded, 
On the other hand, the \textit{Compas Score} pipeline is aimed at predicting the recidivism risk of an individual,   %, and as follows: selection of 9 relevant columns; missing values were deleted; the column \textit{race} was binarized; value transformation of the \textit{label} column for consistency; convertion of \textit{c\_jail\_in} and \textit{c\_jail\_out} columns to days; drop \textit{jail\_in} and \textit{jail\_out} dates and value transformation of column \textit{c\_charge\_degree}.
whereas the goal of the \textit{Census} pipeline is to predict whether annual income for an individual exceeds \$50K. Table \ref{tab:all_pipeline} shows the preprocessing steps for each of these machine learning pipelines. 

\begin{table*}\small%\footnotesize%[]
    \caption{The preprocessing operations included in the machine learning pipelines used in the evaluation.}%\vspace*{-3mm}
     \label{tab:all_pipeline}
    \centering\small
    \begin{tabular}{|c|p{3.2cm}|}
    \multicolumn{2}{c}{\bf German Credit}\\
     \hline
      \textbf{Op} & \textbf{Description}\\
       \hline
       A0 & Value transformation of 13 distinct columns from codes to interpretable terms.\\
       \hline
        A1 & Generation of two new columns from the column \textit{personal\_status}. \\
        \hline
        A2 & The column \textit{personal\_status} was deleted. \\
        \hline
        A3 & 11 categorical columns were OneHot encoded.\\
        \hline
        \end{tabular}
      \hspace{2mm}
     \begin{tabular}{|c|p{3.2cm}|}
     \multicolumn{2}{c}{\bf Compas Score}\\
     \cline{1-2}
      \textbf{Op} & \textbf{Description}\\
       \cline{1-2}
       B0 & Selection of 9 relevant columns.\\
       \hline
       B1 & Missing values were deleted. \\
       \hline
       B2 & The column \textit{race} was binarized. \\
       \hline
       B3 & Value transformation of the \textit{label} column for consistency. \\
       \hline
       B4 & Conversion of \textit{c\_jail\_in} and \textit{c\_jail\_out} columns to days.\\
       \hline
       B5 & Drop \textit{jail\_in} and \textit{jail\_out} dates.\\
       \hline
       B6 & Value transformation of column \textit{c\_charge\_degree}.\\
       \cline{1-2}
        \end{tabular}
       \hspace{2mm}
    \begin{tabular}{|c|p{3.2cm}|}
    \multicolumn{2}{c}{\bf Census}\\
     \hline
      \textbf{Op} & \textbf{Description}\\
       \hline
       C0 & Remove whitespace from 9 columns.\\
       \hline
        C1 & Replace '?' charater for NaN value.\\
        \hline
        C2 & 7 categorical columns were OneHot encoded.\\
        \hline
        C3 & Two columns were binarized.\\ 
        \hline
        C4 & \textit{fnlwgt} column was deleted.\\
        \hline
        \end{tabular}
 \end{table*}

\myparagraph{Capturing provenance.}
%How provenance is captured changes how much information about the underlying processes and data items can be gathered. 
Our work focuses on fine-grained provenance and, as such, it turned out that all provenance queries in Table \ref{tab:provqueries} were answerable.

\begin{figure*}%[tb]
\centering
\includegraphics[width=.8\textwidth]{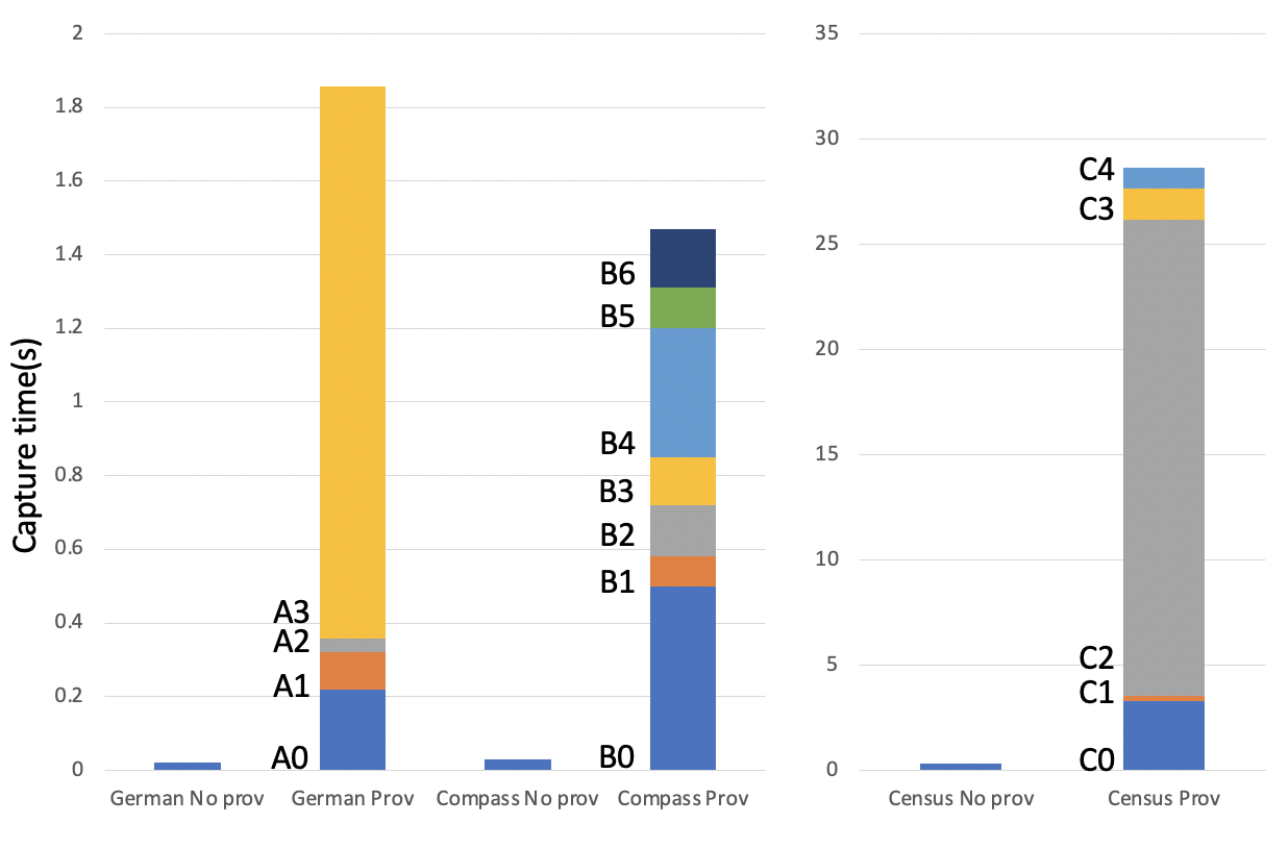}
\caption{Comparison of cumulative provenance capture times, broken down by individual operator.
}\label{fig:allCaptureTimes1}
\end{figure*}

Figure~\ref{fig:allCaptureTimes1} shows the impact of adding provenance capture to a pipeline.  The percentage of overhead of capturing provenance is large compared to executing the system without any provenance at all. However, the actual time to capture the provenance itself is rather low: 1.8s for German Credit, 1.4s for COMPAS, and 28s for Census. These results are 10x faster than the times required for the generation of the same provenance in \cite{chapman2020anonymized}. This improvement is mainly due to the new format used to represent the provenance, and the backend used to store the data, as described in Section~\ref{sec:architecture}. As expected, provenance capture adds computational time to any pipeline execution. However, we note that there are certain complex operations that have a larger impact than others. For instance, in the Census pipeline, the generation of the provenance for operation C2 (One Hot encoding of 7 different columns) requires 22ms. However, this operation introduces 90 new features while the number of records remains unchanged (32,561). Therefore, it generates $32,561\times 90$ new provenance entities. Similarly, operation A3 in the German Credit pipeline is a One-Hot encoding that operates over 11 different columns and creates 38 new features. It follows that this operation creates $1,000\times 38$ new provenance records. Operation B0 in the Compass Score pipeline, which selects 9 columns of data and removes 44 features, is also costly as it generates $7,214\times 44$ provenance records.  Basically, all the other operations generate a limited amount of data provenance and for this reason, they introduce a limited overhead.  The size of provenance generated for the various pipelines is as follows: German Credit 20 MB; Compas Score 71 MB; Census 1.04 GB.

\deleted[id=ac]{For the join operations, we report in Figure %\ref{fig:allCaptureTimes2} 
(The figure has been deleted) the differences between an implementation that makes use of the NumPy library of Python}\footnote{\deleted[id=ac]{\url{https://numpy.org/}}} \deleted[id=ac]{and our optimization based on hash tables, as described in Section \ref{sec:joinImpl}. It turns out that, differently from what happens for the basic implementation, our technique scales very well over the size of the input dataset.}

%\begin{figure}[ht]
%  \includegraphics[width=.8\columnwidth]{figures/joinimprovement.png}
%    \centering
%\caption{Comparison of the implementations of join capture 
%, for each pipeline.
%}\label{fig:allCaptureTimes2}
%\end{figure}

%\subsubsection{Querying Provenance.} 
\myparagraph{Querying Provenance.} 
Provenance would be useless without the ability to query it efficiently. For this, we run all the types of queries reported in Table \ref{tab:provqueries} over the Census dataset, expressing them in Cypher, the query language of Neo4j. Each query was run three times and the resulting time is the average of the three runs. Queries 2 through 6 operate over a single item, a single record, or a single feature, while the others operate over the entire dataset. For the former type of query,  data items, records, and features have been chosen randomly from the output dataset each time the query is run.

\begin{figure*}
    \centering
    \includegraphics[width=.99\textwidth]{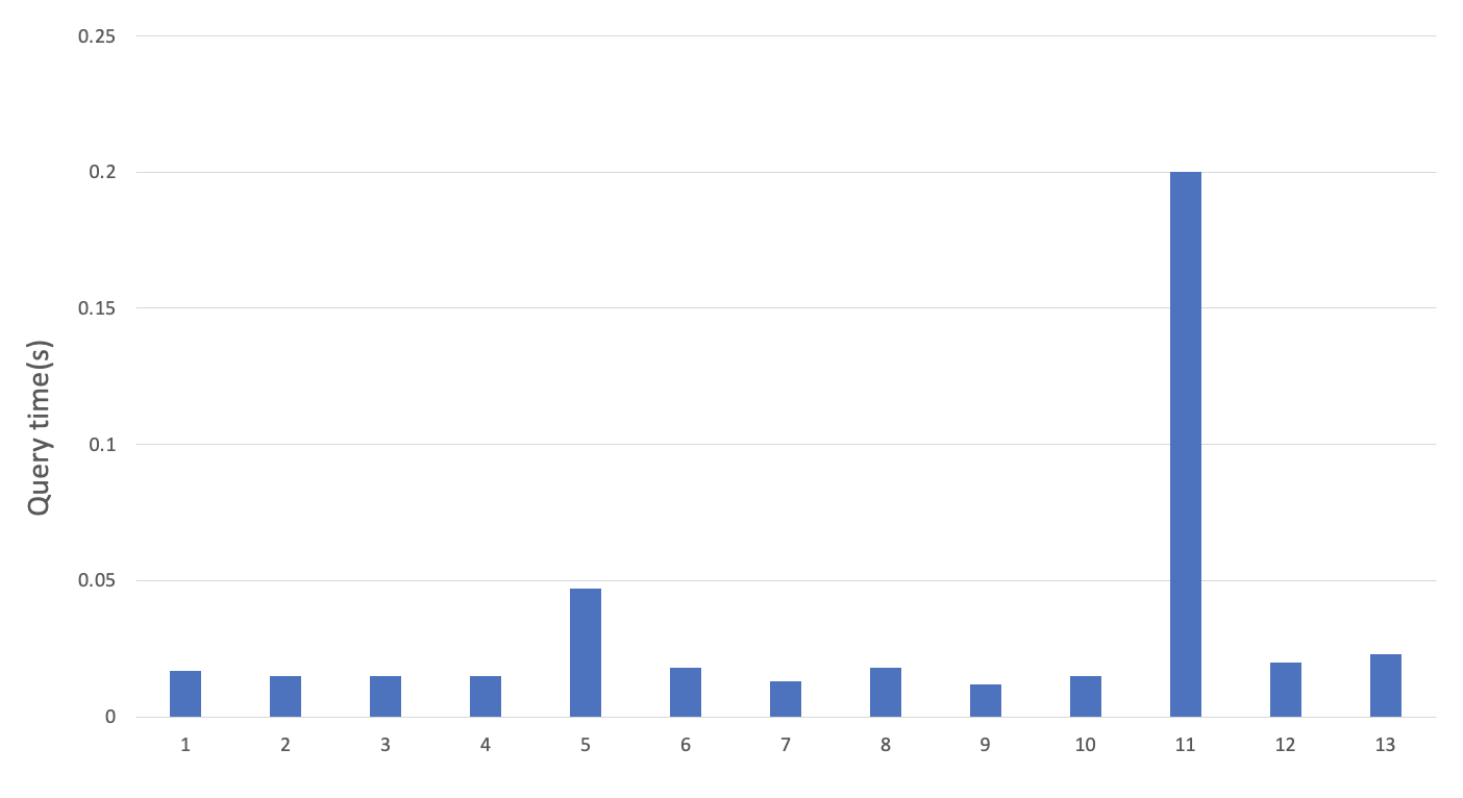}
    \caption{The provenance query times for each type of provenance query are shown in Table \ref{tab:provqueries}.}
    \label{fig:querytime}
\end{figure*}

%\begin{figure}
%    \centering
%    a) \includegraphics[width=.4\columnwidth]{figures/queries1234891011.png} b) \includegraphics[width=.4\columnwidth]{figures/queries56712.png}
%    \caption{The provenance query times for each type of provenance query shown in Table \ref{tab:provqueries}. a) Queries 1, 2, 3, 4, 8, 9, 10, 11; b) Queries 5, 6, 7, 12.}
%   \label{fig:querytime}
%\end{figure}

As shown in Figure \ref{fig:querytime}, the basic provenance queries, particularly those that find or trace paths, are fast. The high-cost queries, such as Query 11, require  additional processing beyond graph traversal. Recall from Section \ref{sec:statement} that queries 10 and 11 traverse the provenance graph \textit{and} search for all future and past derivations of an element. Obviously, depending on the complexity of the operations, the operation can require longer time.

%compute the distribution of the features in the dataset. Query 12 in particular calculates %the distribution of every feature in the dataset before and after every operation. As such, %Query 12 is touching the provenance and every version of every feature of the data. It %represents the worst case scenario for any query.

\subsection{A closer look at operators} \label{sec:tpcdi}

\begin{table*}%[]
    \centering\small
    \caption{Datasets created with the DIGen generator}%\vspace*{-3mm} \label{tab:DIGenData1}
    \begin{tabular}{|c|c|c|c|c|c|} \hline
    \textbf{For Test} & \textbf{Dataset} & \textbf{Scale Factor} & \textbf{Records} & \textbf{Features}  & \textbf{Size} \\ \hline \hline
    \multirow{3}{*}{Basic} & \textbf{Dataset 1} & 3 & 390978 & 45 & 5.2 GB \\ \cline{2-6}
     & \textbf{Dataset 2} & 5 & 650412 & 45 & 8.6 GB \\ \cline{2-6}  & \textbf{Dataset 3} & 9 & 1171107 & 45 & 16 GB \\ \hline
    \multirow{3}{*}{Join} & \textbf{Dataset 4} & 3 & 390978/362342 & 14/5 & 349 MB/117 MB \\ \cline{2-6}
         & \textbf{Dataset 5} & 5 & 650412/602956 & 14/5 & 582 MB/195 MB  \\ \cline{2-6} 
        & \textbf{Dataset 6} & 9 & 1171107/1085239 & 14/5 & 1,05 GB/342 MB \\ \hline
    \multirow{3}{*}{Append} & \textbf{Dataset 7} & 3 & 31581/66689  & 17/17 &  38 MB/81 MB  \\ \cline{2-6}         & \textbf{Dataset 8} & 5 & 55650/138889 & 17/17 & 68 MB/170 MB  \\ \cline{2-6} 
    & \textbf{Dataset 9} & 9 & 109034/283298 & 17/17 & 134 MB/348 MB \\ \hline
    \end{tabular}
    
    \label{tab:DIGenData1}
\end{table*}

\deleted[id=ac]{The previous experiments look mainly at the performance of the methods for capturing the provenance in real world scenarios but do not test the overall scalability of the approach. To accomplish this, w}We use the TPC-DI benchmark \cite{tpcdi} and used DIGen, the data generator provided by TPC, for creating source data and audit information in order to create a known dataset at a larger scale to characterize the behaviour of the reference implementation across a wider range of operators, including Joins and Appends. Specifically, we have created datasets of increasing sizes as described in Table \ref{tab:DIGenData1}: the datasets 1, 2, and 3 involve the trade fact table and the account dimension table, and have been used to measure the \replaced[id=ac]{effect}{scalability} of unary operators in Table~\ref{tab:provop} (DR, FT, ST, IG, VT). Datasets 4, 5, and 6 involve the trade.txt  and HoldingHistory.txt files and were used to measure the \replaced[id=ac]{effect}{scalability} of the join operator (JO in Table~\ref{tab:provop}). Datasets 7,8,9 involve the FINWIRE files and were used to measure the \replaced[id=ac]{effect}{scalability} of the append operator (AP in Table~\ref{tab:provop}).  %In order to test the time required to capture the provenance when performing data preparation pipelines over these datasets, we have considered the operations described in Table~\ref{tab:provop}.  

\begin{table*}\small%[hbt]
\caption{The operations performed on the TCI-DI datasets to test each provenance template.}%\vspace*{-3mm}
\label{tab:provop}
     \centering
     \begin{tabular}{|c|c|p{7cm}|}
        \hline
         \textbf{Op. ID} & \textbf{Operation} & \textbf{Description} \\
          \hline \hline
          DR & Dimensionality Reduction & A column ($D_{*j}$) is removed from the initial dataset.\\ 
          \hline
          FT & Feature Transformation & Transformation on \textit{C\_GNDR} column. Values of gender column are corrected.\\ 
          \hline
          I & Imputation & Imputation on \textit{T\_COMM} column. Null values of trade price column are filled with the average value of the column.\\
          \hline
          ST & Space Transformation & A new column with boolean values is added. 0 if commission value is null, 1 otherwise.\\ 
          \hline
          IG & Instance Generation & Generation of one new record.\\
          \hline
          VT & Value Transformation & Value transformation on \textit{C\_DOB} column. Invalid date of birth are replaced with NaN values. \\
          \hline
          JO & JOin &  Left outer join between Trade table and Holding History on Trade ID \\
          \hline
          AP & APpend & FINWIRE files from 1967 to 1993  and from 1994 to 2017 were concatenated and then used to perform an append   \\
          \hline

     \end{tabular}
 \end{table*}
 
 \begin{figure*}%[hbt]
    \centering
    \includegraphics[width=.99\textwidth]{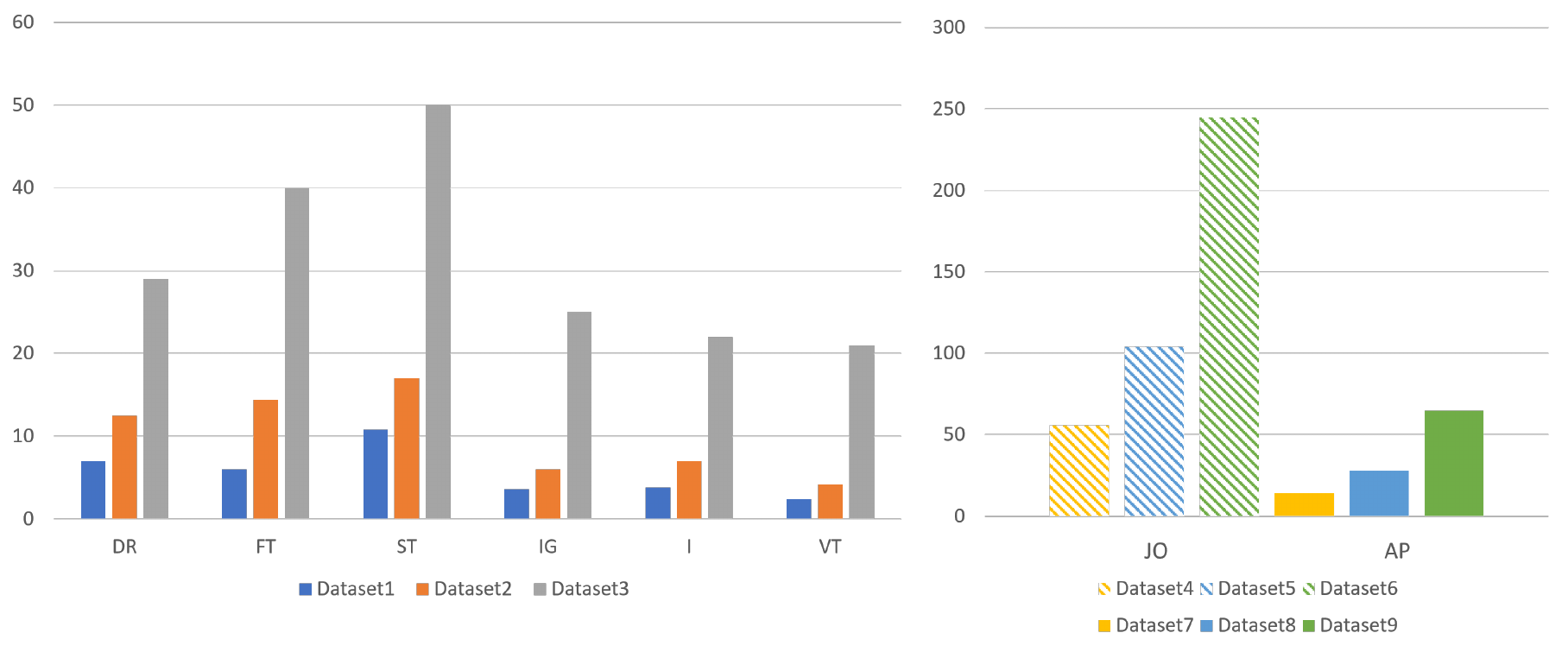}
    
\vspace{3mm}
\small
     \begin{tabular}{|c|c|c|c|}
        \hline
        \textbf{Operation} & \textbf{Dataset 1} & \textbf{Dataset 2} & \textbf{Dataset 3} \\
         \hline \hline 
        Dimensionality Reduction & 10 MB & 17 MB & 31 MB \\
        \hline
        Feature Transformation & 70 MB & 120 MB & 218 MB \\
        \hline
        Imputation & 15 MB & 25 MB & 46 MB \\ 
        \hline
        Space Transformation & 60 MB & 101 MB & 183 MB \\ 
        \hline
        Instance Generation & 10 MB & 17 MB & 31 MB \\ 
        \hline
        Value Transformation & 110 KB & 400 KB & 550 KB \\ 
        \hline
        Join & 1.02 GB & 1,78 GB & 3.04 GB \\ 
        \hline
        Append & 390 MB & 779 MB & 1,6 GB \\ 
        \hline
     \end{tabular}

\caption{Capture time (in seconds) and storage space for each operation} \label{fig:capture_time}
\end{figure*}
 
Figure \ref{fig:capture_time} shows how long and how much space it takes to capture and record provenance for each operation. 
The capture mechanism scales rather well with the size of the dataset and it turns out that pre-processing operations that only affect a small number of data values, such as Instance Generation (IG), are fast.  
Value Transform (VT) and Imputation (I), in this particular evaluation setup, are also fast as they only operate over a small number of items. 
On the other hand, the operations that generate more provenance, such as Feature Transformation (FT), Space Transformation (ST), and Dimensionality Reduction (DR), take more time. In particular, ST needs to create provenance data for every new value in the new column. Join (JO) and Append (AP) operations require more time as they need to generate a quite large quantity of provenance. In addition, JO is more costly as it operates over two input tables.

\begin{figure*}%[hbt]
    \centering
    \includegraphics[width=.8\textwidth]{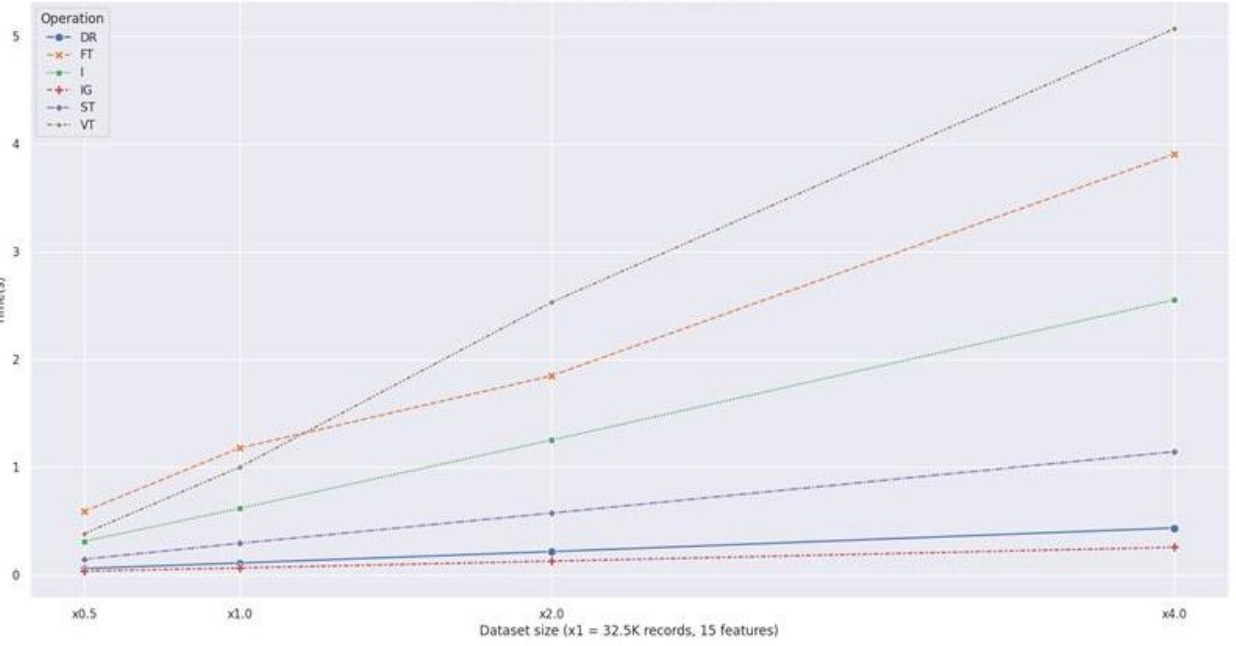}

    \includegraphics[width=.8\textwidth]{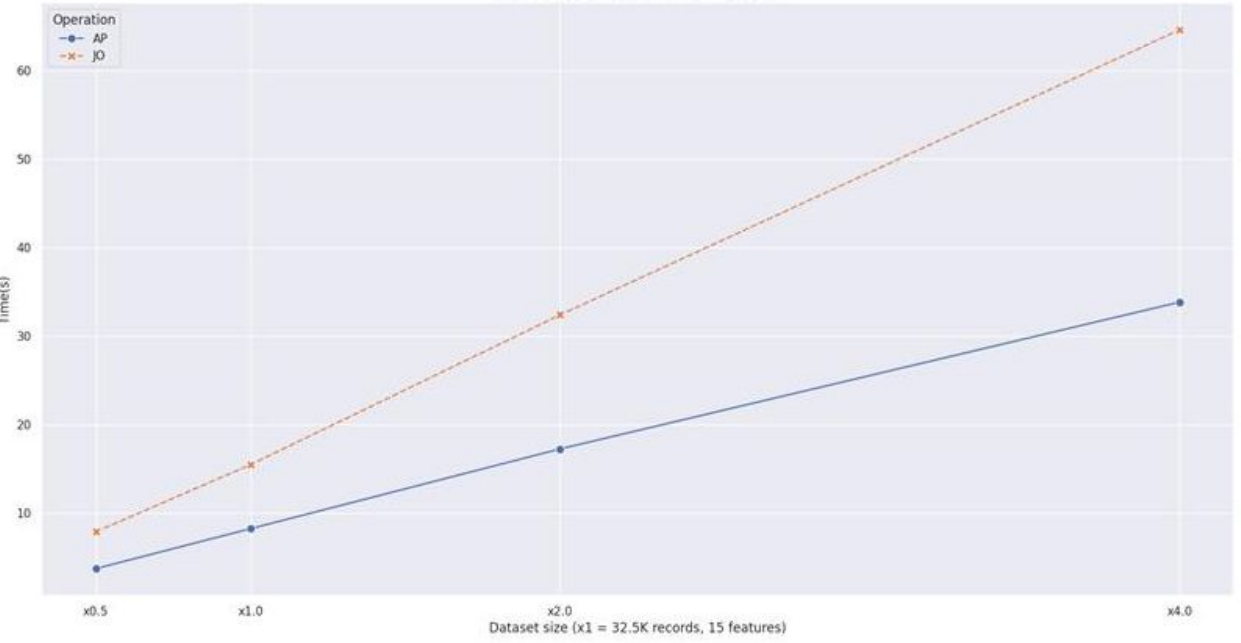}
\caption{Capture time (in seconds) by varying the size of a fixed dataset} \label{fig:capture_time2}
\end{figure*}

The fact that provenance capture scales gracefully with the dataset size is confirmed by another experiment, whose results are reported in Figure \ref{fig:capture_time2}, in which we have executed the same types of operations in Table \ref{tab:provop} by just varying the number of records of a fixed input (the Census dataset). Since the evaluation setup here is different, the various operators exhibit a different behavior in terms of relative performances, but their computational time remains quite low and grows linearly with the dataset size.

\subsection{Use Case Analysis}\label{sec:dsseanalysis}

Table \ref{tab:realprovqueries} contains a collection of real-world scenarios in which data scientists try to understand what is happening within a machine-learning pipeline. These use cases have been gathered from the Data Science Stack Exchange\footnote{\url{https://datascience.stackexchange.com/}} (DSSE) by selecting questions about the construction of a data preparation pipeline using the Orange framework. The provenance queries that can provide support to these issues refer to those in Table~\ref{tab:provqueries} (Page~\pageref{tab:provqueries}), in which, for each query, it is reported the input data and the expected output that can help the developer to debug the pipeline.

%In Appendix \ref{app:usecases}, we report a selection of real questions of data scientists taken from the Data Science Stack Exchange\footnote{\url{https://datascience.stackexchange.com/}} for which the analysis of data provenance can provide an answer. 

\begin{table*}%[hbt!]
\caption{Issues identified in real Machine Learning pipelines and provenance queries that can provide support to them.}\label{tab:realprovqueries}
\centering\small
\begin{tabular}{|c|p{10cm}|p{2cm}|}
\hline
\textbf{Id} & \textbf{Data Science Stack Exchange Use Cases}  & \textbf{Prov. query id}\\\hline \hline
UC1 & When applying the Predictions widget on the same training dataset, the results (i.e. probability scores) are different: \url{https://datascience.stackexchange.com/questions/32382/orange-predictions-widget-on-same-data-gives-different-results} & PQ1\\ \hline
UC2 & Differences in the predictions and goodness-of-fit of R2 metric for the linear regression model on Orange and Scikit-learn: \url{https://datascience.stackexchange.com/questions/32678/orange-linear-regression-and-scikit-learn-linear-regression-gives-different-resu} & PQ2 \\ \hline
UC3 & After performing image classification using an ML model, prediction probabilities are constant on test images \url{https://datascience.stackexchange.com/questions/38320/orange3-image-classification} & PQ3\\ \hline
UC4 & From a constructed workflow using image classification (add-on widgets) ascertain whether the workflow performs transfer learning:  \url{https://datascience.stackexchange.com/questions/19240/using-orange3-to-predict-image-class} & PQ3 \\\hline
UC5 & Application of the Test and Score and Predictions widget on the same data utilising the same ML model; produces differing results: \url{https://datascience.stackexchange.com/questions/20572/why-orange-predictions-and-test-score-produce-different-results-on-the-sam} & PQ3 \\ \hline
UC6 & When applying the Impute widget during preprocessing on the train/test dataset, the same values are predicted for all rows: \url{https://datascience.stackexchange.com//questions/15264/orange-3-same-prediction-for-all-of-my-data-when-using-impute-widget}  & PQ4, PQ5, PQ6, PQ11, PQ12\\ \hline
UC7 & Inaccuracy in the prediction of target variable using k-NN and linear regression ML models in an Orange workflow: \url{https://datascience.stackexchange.com/questions/36537/how-to-properly-predict-date-using-orange-3} & PQ7, PQ8, PQ9, PQ10 \\ \hline
UC8 & Disproportionate allocation of labels after performing data analysis and modeling (inaccurate classification accuracy): \url{https://datascience.stackexchange.com/questions/37471/dataset-with-disproportionately-more-of-a-single-label-than-any-other} & PQ11, PQ12\\ \hline
\end{tabular}
\end{table*}

To highlight how the fine-grained provenance captured with our approach can be used to answer one of these questions consider, for instance, the UC8 use case. 
In this scenario, the user is struggling with an incorrect high accuracy of a model. Ultimately, this is because of an imbalanced input dataset. Using the Provenance Query \textit{Impact on Feature Spread} from Table \ref{tab:provqueries} on the input dataset, it is possible to identify the change of feature spread after a pre-processing operator that rebalances the dataset. 

\subsection{Provenance Exploration}\label{sec:provexp}

Unlike many provenance systems, which focus on the presentation and navigation of the provenance graph, we have developed a tool\footnote{The code of this tool is publicly available on GitHub.} in which the provenance graph is mainly used as a backbone to explore and identify problems within the pipeline through a user-friendly interface,  which does not require the specification of complex queries over the data provenance.
This is done by automatically extracting, from the provenance and other metadata, useful information on the changes operated by the individual operations on the input dataset(s).

\begin{figure}
     \includegraphics[width=.6\columnwidth]{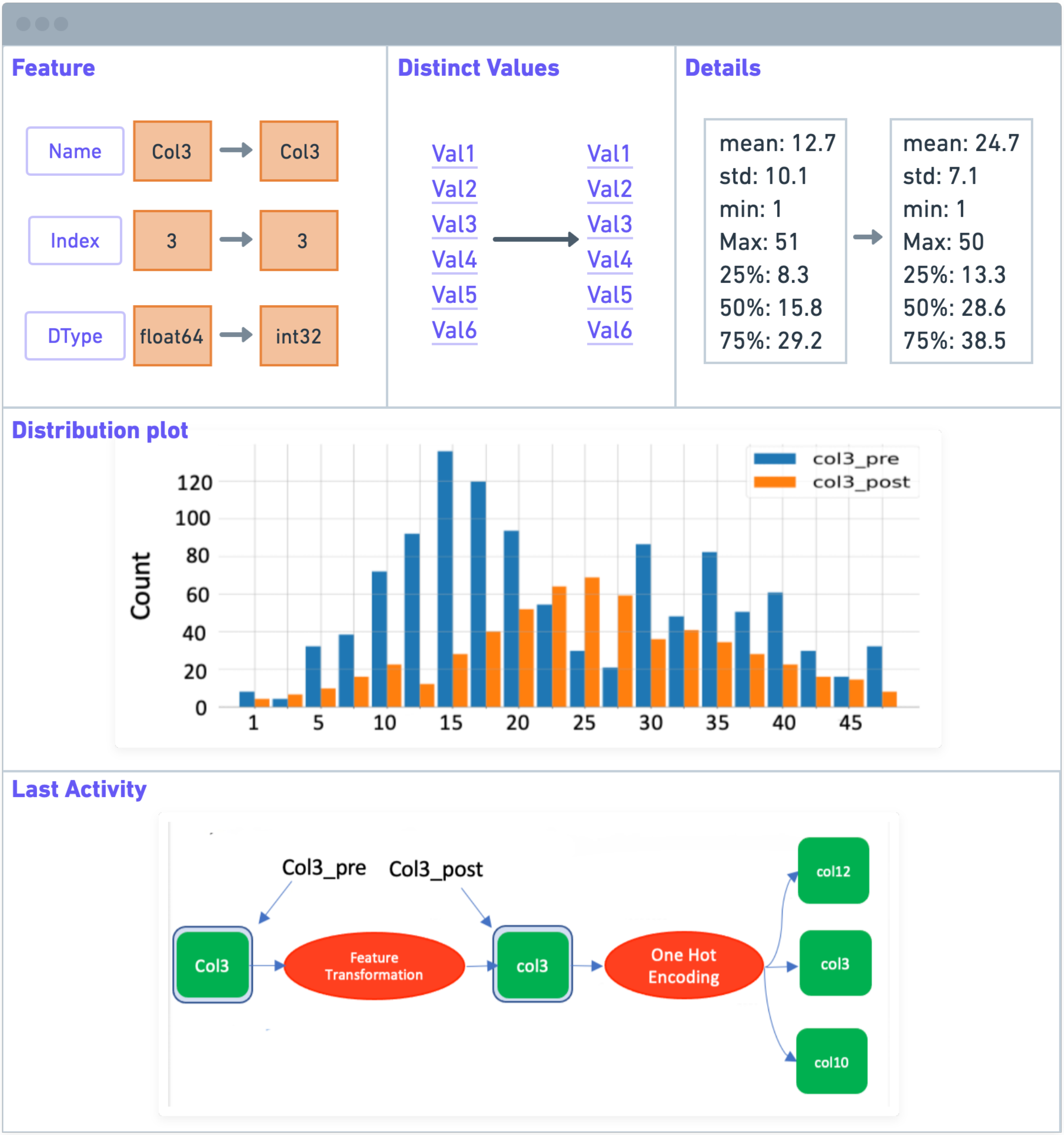}
    \centering
\caption{Example of a GUI showing how data changes after a transformation process that operates locally, at the feature level.}
\label{fig:interface}
\end{figure}

\begin{figure}
    \centering
     \includegraphics[width=.6\columnwidth]{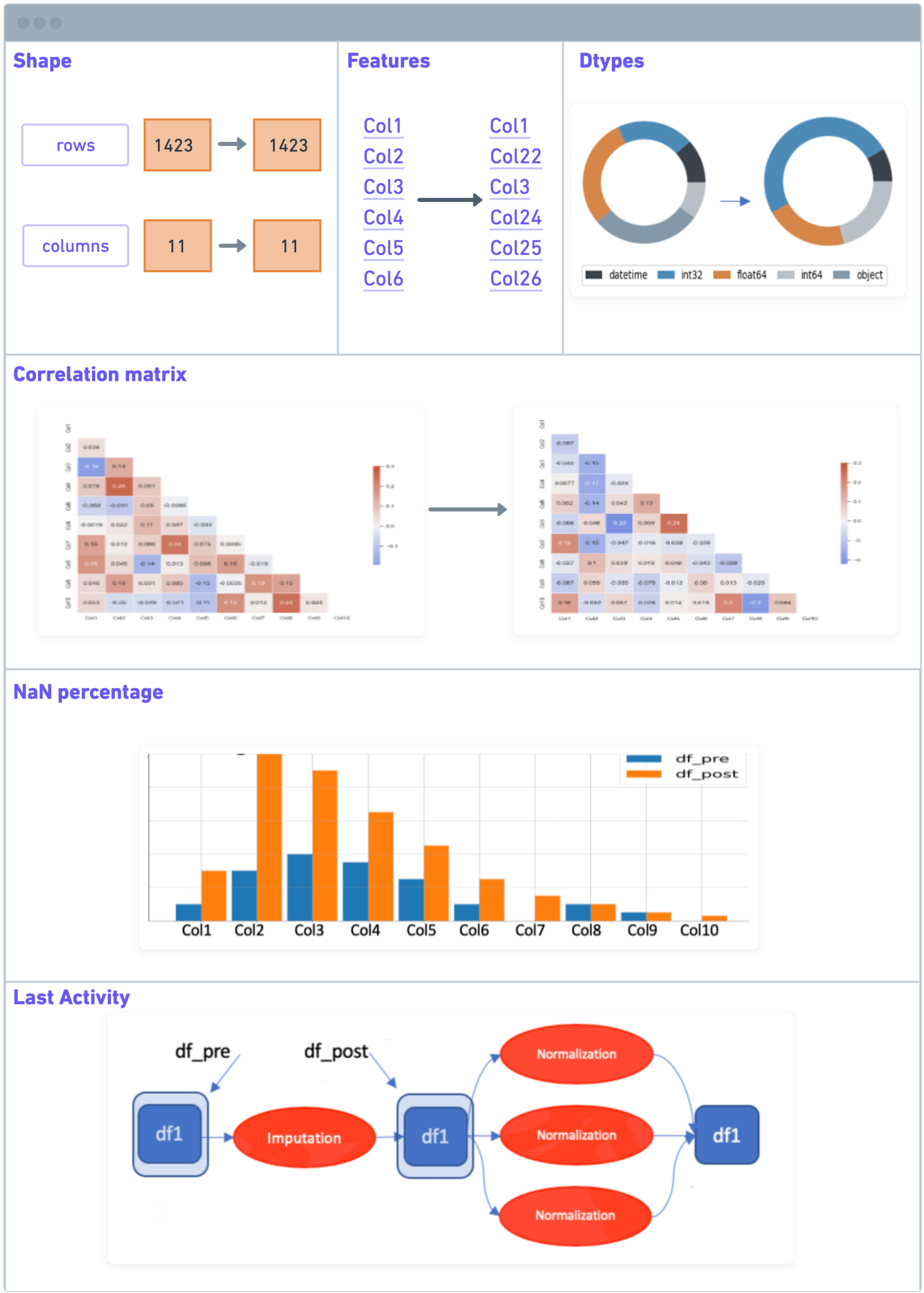}
    \caption{Example of a GUI showing how data changes at the dataset level, after an imputation operation over several columns.}
    \label{fig:interface2}
\end{figure}

An example of this kind of interaction is shown in the GUIs of our tool reported in Figures~\ref{fig:interface} and~\ref{fig:interface2} respectively: basically, depending on the type of operator that was applied, the data scientist can ``zoom in'' to a transformation of interest (bottom of the figures) and inspect the ``before/after'' effect of its execution either at the level of values within a column, in the case of a \textit{local} transformation, or at the level of the entire dataset, in the case of a \textit{global} transformation. 

A local type of transformation is illustrated in Figure~\ref{fig:interface}, where a data transformation operation, which modifies values in \textsf{col3}, is represented in the provenance fragment at the bottom. The user is then able to navigate through the retrieved provenance, identify the pre- and post- states for \textsf{col3}, and visualise their differences in terms of summary statistics (top right), values distribution (center), and optionally each value can be inspected (top middle).

In contrast, Figure~ \ref{fig:interface2} shows the effect of an imputation step that operates globally. As this may change more than one column at a time (for instance, using Multiple Inference), here the GUI displays salient differences at the dataset level. We can see for instance that the operation has not changed the number of rows and columns of the dataset (top left), but the imputation has updated the content of several columns (col2, col4, col5, col6, see top middle), and has altered the percentage of null values (bar chart). We can also see the changes in the correlation between each pair of columns in the dataset before and after the operation is performed.

\subsection{Comparison to other provenance collection systems} \label{sec:comparison}
There are many provenance systems that can be deployed to capture provenance of workflow-like executions. In this section, we look at some of the main players and compare them to the provenance in this work. Because the implementation in this work was a reference implementation for exploration, not production deployment, we feel that an execution benchmark between the systems is uninteresting.

\begin{table*}[t]
    \centering\small
        \caption{Comparison of this work with other provenance capture systems}
    \label{tab:systemCompare}
    \begin{tabular}{|p{2cm}|p{3cm}|p{5.8cm}|p{1.8cm}|} 
    \hline
    \textbf{Provenance System} & \textbf{Deployed on}  & \textbf{Capture Method} &  \textbf{Granularity} \\ \hline
        This work & Jupyter Notebooks with dataframes  & Operator derived from comparison of before-after dataframes; provenance templates associated with operator.  & attribute \\ \hline
        MLInspect \cite{grafberger2021}\cite{grafberger2022}\cite{grafberger2023provenance} &  dataframes & Python's inspect module and Monkey patch to identify operators at execution to build a DAG of operators & tuple \\ \hline
        Perm \cite{glavic2009perm,Glavic2009} & Relational databases  & Query rewrite to add and propagate provenance attributes to the output of the original query. & tuple\\ \hline
        Vamsa \cite{namaki2020vamsa} & python scrips with dataframes & Static analysis of script to identify inputs, parameters and libraries; knowledge base to provide semantic meaning. & dataset \\ \hline
    \end{tabular}
\end{table*}

\textit{Perm \cite{glavic2009sig,Glavic2009}.}  Perm uses query rewrites to add and propagate provenance attributes to the output of the original query. It can capture and propagate provenance for ASPJ queries and set operations. It is implemented on PostgreSQL and tested against TPC-H with an overhead on TPC-H queries of 3-4x. The queries outside of this norm include very complex queries with aggregation, such as an aggregation over a join on 8 tables with a grouping on a functional expression. Perm allows lazy and eager computation of provenance, SQL query facilities, and support for external provenance. Perm outperformed previous approaches by a factor of 30. While the execution of Perm is impressive, it fundamentally relies on a technology that is not appropriate for the problem within notebooks focused on in this work. 

\textit{MLInspect \cite{grafberger2021,grafberger2022}.} The MLInspect system uses Python’s inspect module or Monkey patch to identify function calls within python scripts and build a DAG of relationships and interactions in the pre-processing pipeline. This run-time representation is updated as the developer changes the scripts based on the standard dataframe operators. A user can annotate tuples, and specify the inspections that need to occur (e.g. inspect for statistical parity of protected group). As the script is executed, this DAG is stepped through, and the operator is passed for inspection.  The focus of the MLInspect is to analyze data distribution after operators based on the pre-specified inspections. Provenance at the tuple level is supported through the adaptation of user annotations by recording the pre-assigned tuple-id and the operator applied. Our work has a very different focus, and as such the provenance requirements are different. In MLInspect, provenance can be added at the tuple level and used to support data distribution change analysis; our work provides a much finer-grained provenance at the attribute level allowing for debugging specific value changes.

\textit{Vamsa \cite{namaki2020vamsa}.} The Vamsa system uses static analysis to build a syntax tree to identify inputs, parameters and libraries. It then uses a knowledge base to provide semantic meaning to these items and log it in the provenance. However, Vamsa relies upon a pre-populated knowledge base which maps the set of functions identified in the code to semantic "operators". This approach restricts provenance generation to known operators. The Vamsa experimentation shows that their coverage ranges between  74.48\% - 97.08\%. Our approach relies on inspection of the dataframe before and after to identify the type of transformation that occurred (e.g. horizontal reduction) instead of relying upon a pre-created knowledge base. In addition to this collection difference, Vamsa identifies and store provenance information at the dataset level. For instance, it will identify entire rows retained or dropped; it can identify whether a column within a dataframe is used. However, it does not track individual changes to attributes. While these can be later derived by understanding what operators were applied to which rows or columns, this information is not innately stored. A combination of Vamsa and this work would be interesting future work, in which Vamsa is used to identify the pre-stocked operators, and for the remainder, DPDS identifies what is happening in the via dataframe changes and templates of provenance.

\section{Related Work} \label{sec:relatedwork}

This paper substantially advances 
%our 
previous work~\cite{chapman2020anonymized} by: (i) extending the set of core operations with methods for combining different datasets to any operator that modifies a dataframe, (ii) replacing the manual instrumentation at the script level required by the  analysts with a method for the identification of provenance for most of the operators through dataset change
%moving from gathering provenance of the operator execution to the identification of provenance through dataset change
% replacing the manual instrumentation at the script level required by the  analysts with an automatic method for detecting, at run-time, most of the operators applied to a dataset, 
(iii) adopting a graph-based data management system for storing and querying in an effective and efficient way the collected provenance, (iv) \deleted[id=ac]{increasing the efficiency and scalability of the approach by an order of magnitude through 
%an implementation that includes 
multiple optimizations, and (v) }performing %new 
experiments for empirical validation \added[id=ac]{and qualitative comparison to previous work}.

Established techniques and tools are available to generate provenance, and \textit{provenance polynomials} through query instrumentation. However, these operate in a  relational database setting and assume that queries use relational operators~\cite{NiuKGGLR17,ArabFGLNZ17,Glavic2009}.
While we show how some of the pipeline operators considered in this work map to  relational algebra, this is not true for all of them, so we prefer to avoid  techniques that are tightly linked to SQL or to first-order queries~\cite{LeeKLG17} as these would preclude other types of operators from being included in the future. 
We, therefore, consider this an unwise strategy in an ``open world'' of data pre-processing operators, consider e.g. \textit{one-hot} and  other kinds of categorical data encodings. 
We also note that tools that operate on a database back-end,  like \textit{GProm}~\cite{NiuKGGLR17}, \textit{Smoke}~\cite{psallidas2018smoke} and older ones like \textit{Post-it} \cite{ChiticariuTV05}  for provenance capture cannot be used in our setting.
Interestingly, extensions to the polynomials approach have been proposed to describe the provenance of certain linear algebra operations, such as matrix decomposition and tensor-product construction~\cite{YanTI16}. While these can potentially be useful, it is a partially developed theory with limited and specialised applicability. 

Moving beyond relational data provenance, capturing provenance within scripts is also not new, but efforts have mostly focused on the provenance of script  definition, deployment,  and execution~\cite{pimentel2019survey}. Specifically, a number of tools are available to help developers build machine learning pipelines~\cite{agrawal2019,demvsar2013orange,shang2019} or debug them~\cite{vartak2018mistique}, but these lack the ability to explain the provenance of a certain data item in the processed dataset. Others link provenance to explainability in a distributed machine learning setting~\cite{scherzinger2019best} but without offering specific tools.
Amazon identifies that there are common and reusable components to a machine learning pipeline, but that there is no way to track the exploration of pipeline construction effectively, and calls for metadata capture to support reasoning over pipeline design \cite{Schelter2018DeclarativeMM}. 
Vamsa \cite{namaki2020vamsa} attempts to tackle some of these problems by gathering the provenance of pipeline design. However, the resulting provenance documents contain information such as the invocation of specific ML libraries, by way of automated script analysis, rather than data derivations. 
Some systems are designed to help debug ML pipelines. BugDoc \cite{bugdoc2020} looks at changes in a  pre-processing pipeline that cause the models to fail, where high-level script and orders are used to identify bad configurations.  Others provide quality assurance frameworks~\cite{studer2020towards} or embedded simulators to estimate the fairness impacts of a particular pipeline \cite{damour2020}. 
Again, however, these are not geared for deep data introspection. Priu~\cite{WuTD20}, helps users understand data changes, particularly deletions, that are used in regression models. Unfortunately, this work only tracks deletions and not additions or updates to data. 

Recently, \cite{grafberger2022data} have utilized provenance to understand the changes in data distribution in the ML pipeline using predefined ``inspections'' that look at the data at specific operators within the pipeline, which supports the reason for undertaking this work and which we expand by unobtrusively capturing provenance from any operator. Meanwhile, \cite{rupprecht2020} combines system level provenance information with application-level log files to recreate the provenance of data science pipelines without impacting the pipeline developer.
Other tools record the execution of generic (python) scripts, but fail to capture detailed data provenance, like NoWorkflow \cite{pimentel2016fine,pimentel2017noworkflow}. This has been combined with YesWorkflow \cite{mcphillips2015yesworkflow,zhang2017using} which provides a workflow-like description of scripts, but again without a focus on data derivations.  

A further class of tools instrument scripts that are specifically designed for Big Data processing frameworks:
\cite{ikeda2012provenance} (Hadoop), \cite{guedes2018,tang2019,interlandi2016,psallidas2018smoke} (Spark).
They provide detailed information mostly for debugging purposes but are restricted in their scope of applicability.

%\acnote{Someone read this. Long-winded and weak. Fixes and suggestions appreciated.}
Recently, a method for fine-grained provenance capture that is application-agnostic has been proposed \cite{rupprecht2020}. Here, provenance from the low-level OS through to high-level application-specific logs is merged to create a provenance record that contains the maximum information available for the minimum impact on developers. However, it is not obvious what fine-grain provenance can be extracted from such an approach, while our work provides a firm basis for the provenance information that should be captured. Interesting future work includes determining how much of the provenance we specify can be collected by  \cite{rupprecht2020}.

Finally, the method proposed within Section \ref{sec:algorithms} in which the change of the data is observed instead of the operator is similar to techniques discussed in \cite{blount2021observed}. While Blount describes the general setup of inferring the provenance record based on identified changes in the data, our work provides a functioning implementation for a large class of operators.

\section{Conclusions and Future Work} \label{sec:concl}

% \added[id=pm]{ADD NOTE ABOUT ``OPENING THE BLACK BOX'' AND INTERACTIVE / MARKUP BASED ANNOTATIONS TO MAKE PROVENANCE MORE PRECISE}

% \pmnote{one of the possible future directions: a formal framwrok to describe generla provenance from workflows / scripts that aligns with or complements eg prov polynomials.}

In this work, we focus on fine-grained data provenance for machine learning pipelines irrespective of the pipeline tool used. Because a substantial effort goes into selecting and preparing data for use in modelling, and because changes made during preparation can affect the ultimate model, it is important to be able to trace what is happening to the data at a fine-grain level. 

We highlight several real use cases to motivate the need for fine-grained provenance from the Data Science Stack Exchange (DSSE)\footnotemark[1]. We identify the classic provenance queries that are needed to provide information to answer these use cases. We then identify a set of provenance templates that can be deployed across a set of machine learning pipeline operators and implement them. 

We depart significantly in this work from previous implementations within python and ML environments, by using \textit{observed} changes in the data to determine the provenance. 
Based on observations of the changes between dataframes, we choose the appropriate template for provenance generation. We have tested our implementation over real-world ML benchmark pipelines for utility and basic performance \added[id=ac]{with both classic ML pipelines and TCP-DI}. \deleted[id=ac]{In order to investigate scalability issues with our design, we also use the TCP-DI generator and apply several operators over that data at scale.} Our results indicate that we can collect fine-grained provenance that is both useful and performant.

Future investigation into optimization techniques that aim at reducing the provenance data, using composite generation, to the minimum that is needed to support given provenance queries, as well as methods for taking advantage of collected provenance data to support the design of new pipelines is required to continue making provenance more efficient and useful. \added[id=ac]{This work looks expressly at the pre-processing tools leading up to the machine learning black box, thus it does not track provenance models for the trained data, e.g. between predictions and training data. However, this work has been used by \cite{pina2023} to create an entire tracking of data from pre-processing through deep learning. Future work in this area includes understanding the granularity of provenance required for users of deep learning systems.}

\bibliographystyle{unsrt}  
\bibliography{prov}

\end{document}